\documentclass[12pt, twoside]{article}\usepackage[]{graphicx}\usepackage[]{color}
\makeatletter
\def\maxwidth{ %
  \ifdim\Gin@nat@width>\linewidth
    \linewidth
  \else
    \Gin@nat@width
  \fi
}
\makeatother

\definecolor{fgcolor}{rgb}{0.345, 0.345, 0.345}

\usepackage{framed}
\makeatletter
 {\par\unskip\endMakeFramed%
 \at@end@of@kframe}
\makeatother

\definecolor{shadecolor}{rgb}{.97, .97, .97}
\definecolor{messagecolor}{rgb}{0, 0, 0}
\definecolor{warningcolor}{rgb}{1, 0, 1}
\definecolor{errorcolor}{rgb}{1, 0, 0}
\newenvironment{knitrout}{}{} 

\usepackage{alltt}

\usepackage{multirow}

\definecolor{dkrd}{rgb}{.851,.373,.008}
\definecolor{dkgr}{rgb}{.106,.62,.467}
\definecolor{dkpu}{rgb}{.459,.439,.702}

\usepackage[margin=1in]{geometry}

\usepackage[font=doublespacing]{caption}
\usepackage{setspace}

\usepackage{amsthm, amssymb, bm, dsfont}
\usepackage[leqno]{amsmath}
\usepackage{bigints}
\usepackage{apptools}

\usepackage{mathrsfs}
\DeclareFontFamily{U}{calligra}{}
\DeclareFontShape{U}{calligra}{m}{n}{<->callig15}{}

\usepackage{booktabs, makecell, enumitem, tabularx}
\newcommand{\ra}[1]{\renewcommand{\arraystretch}{#1}}

\newcolumntype{Y}{>{\centering\arraybackslash}X}

\usepackage{xparse}

\usepackage{xr}
\usepackage[natbibapa]{apacite}
\usepackage{etoolbox}

\AtBeginEnvironment{APACrefURL}{\renewcommand{\url}[1]{}}
\RequirePackage[colorlinks,linkcolor=blue,citecolor=blue,urlcolor=blue]{hyperref}
\usepackage{cleveref}

\binoppenalty=\maxdimen
\relpenalty=\maxdimen

\usepackage{placeins}
\usepackage[nolists, nomarkers, noheads]{endfloat}

\def\equationautorefname~#1\null{%
  (#1)\null
}

\allowdisplaybreaks

\makeatletter
\newcommand\Autoref[1]{\@first@ref#1,@}
\def\@throw@dot#1.#2@{#1}
\def\@set@refname#1{
    \edef\@tmp{\getrefbykeydefault{#1}{anchor}{}}%
    \xdef\@tmp{\expandafter\@throw@dot\@tmp.@}%
    \ltx@IfUndefined{\@tmp autorefnameplural}%
         {\def\@refname{\@nameuse{\@tmp autorefname}s}}%
         {\def\@refname{\@nameuse{\@tmp autorefnameplural}}}%
}
\def\@first@ref#1,#2{%
  \ifx#2@\autoref{#1}\let\@nextref\@gobble
  \else%
    (\ref{#1}
    \let\@nextref\@next@ref
  \fi%
  \@nextref#2%
}
\def\@next@ref#1,#2{%
   \ifx#2@,~\ref{#1}\let\@nextref\@gobble
   \else, \ref{#1}
   \fi%
   \@nextref#2%
   )
}
\makeatother

\usepackage{autonum}

\crefformat{equation}{(#2#1#3)}
\crefrangeformat{equation}{(#3#1#4) to~(#5#2#6)}

\AfterEndPreamble{%
\let\autoref\cref
}%

\makeatletter
\newcommand{\restore@Environment}[1]{%
  \AtBeginDocument{%
    \csletcs{#1*}{#1}%
    \csletcs{end#1*}{end#1}%
  }%
}
\forcsvlist\restore@Environment{alignat,equation,gather,multline,flalign,align}
\makeatother

  \renewcommand{\vec}[1]{\boldsymbol{#1}}
  \renewcommand{\v}[1]{\vec{#1}}
  \newcommand{\norm}[1]{\left\Vert#1\right\Vert}
  \newcommand{\abs}[1]{\left\vert#1\right\vert}

  \DeclareMathOperator{\logit}{logit}

  \newcommand{\paren}[1]{{\left(#1\right)}}
  \newcommand{\brc}[1]{{\left\{#1\right\}}}
  \newcommand{\brk}[1]{{\left[#1\right]}}
  \newcommand{\inv}[1]{{#1^{-1}}}
  \newcommand{\given}[2]{{\left.#1\right\vert#2}}

  \newcommand{\Cov}[2]{{\text{Cov}\left(#1, #2\right)}}

  \NewDocumentCommand\distn{mg}{
    #1\IfNoValueTF{#2}{}{\left(#2\right)}
  }

  \newcommand{\vs}{{\v s}}
  \newcommand{\vx}{{\v x}}
  
  \newcommand{\D}{{\mathcal D}}
  \newcommand{\DS}{{\mathcal S}}
  \newcommand*{\nolink}[1]{%
    \begin{NoHyper}#1\end{NoHyper}%
  }

\title{Approximate Bayesian Inference via Sparse grid Quadrature Evaluation for Hierarchical Models}

\usepackage{fancyhdr}
\usepackage{authblk}
\author{Joshua Hewitt}
\author{Jennifer A. Hoeting}
\affil{Colorado State University}

\date{\vspace{-5ex}}

\IfFileExists{upquote.sty}{\usepackage{upquote}}{}
\begin{document}

\singlespace

\maketitle

\doublespace

\begin{abstract}
{\textbf{Abstract:}}

We combine conditioning techniques with sparse grid quadrature rules to develop
a computationally efficient method to approximate marginal, but not necessarily
univariate, posterior quantities, yielding approximate Bayesian inference via
Sparse grid Quadrature Evaluation (BISQuE) for hierarchical models.  BISQuE
reformulates posterior quantities as weighted integrals of conditional
quantities, such as densities and expectations.  Sparse grid quadrature rules
allow computationally efficient approximation of high dimensional integrals,
which appear in hierarchical models with many hyperparameters.  BISQuE reduces
computational effort relative to standard, Markov chain Monte Carlo methods by
at least two orders of magnitude on several applied and illustrative models.
We also briefly discuss using BISQuE to apply Integrated Nested Laplace
Approximations (INLA) to models with more hyperparameters than is currently
practical.

\end{abstract}


\section{Introduction}
\label{sec:intro}

Computationally efficient posterior approximation remains a key challenge
in applied \linebreak Bayesian
analyses, especially for hierarchical models.  Hierarchical Bayesian models
allow flexible modeling of complex data, but make posterior inference
challenging because simple, conjugate distributions are typically unavailable.
Posterior densities, expectations, and other quantities involve computing
integrals that often require numerical approximation.  The required
approximations can be computationally expensive or challenging since many
hierarchical models include many unknown parameters, thus integrals are defined
over high dimensional state spaces.  Sampling-based approaches, like
Markov chain Monte Carlo (MCMC) methods, are widely used because they are
generally reliable and relatively simple to implement \citep{Gelfand1990}.
However, MCMC approximations can be computationally expensive for many models.
Full conditional posterior distributions required by a Gibbs sampler can be
difficult to sample efficiently or lead to highly correlated Monte Carlo
samplers.  As a result, if $n$ dependent samples are drawn via MCMC
methods, the stochastic approximation error rate can often be higher than the
error $\mathcal O_p\paren{n^{-1/2}}$ for direct Monte Carlo approximations.
Alternate approaches for Bayesian approximation are available via a range of
stochastic and deterministic methods, including Laplace and Integrated Nested
Laplace approximations \citep{Tierney1986, Rue2009a}, classical
quadrature-based approximations \citep{Naylor1982}, Variational Bayes
\citep{Attias2000}, and Approximate Bayesian Computing
\citep{Tavare1997, Rubin1984}.  Generally, each method is
motivated by computational issues and structures found in different classes of
models, so no method is necessarily well-suited for all hierarchical models.
In particular, technical limitations of Integrated Nested Laplace
approximations (INLA) and classical quadrature motivate us to develop a
strategy to yield approximate Bayesian Inference via Sparse grid Quadrature
Evaluation (BISQuE) for a wider range of hierarchical models.

INLA approximates marginal posterior distributions by using a discrete numerical
integration grid of hyperparameters to average over Laplace approximations of
conditional posterior densities.  The method is developed for models that link
observations to latent Gaussian variables through link functions, similar to
generalized linear models.  The INLA approximation enables fast inference for a
wide range of scientifically relevant models.  However, it can sometimes be
difficult to reformulate models to have the latent
Gaussian structure required by the INLA framework.  Additionally, the numerical
integration in INLA can become computationally infeasible for models with
many hyperparameters.  The latter issue is a limitation shared by classical
quadrature-based approximations for posterior quantities.

Classical quadrature methods can approximate marginal posterior
distributions and expectations for general Bayesian models, but like INLA, the
models must have relatively small dimension \citep{Naylor1982}.
Quadrature methods approximate an integral by evaluating its integrand at
deterministic nodes and weighting the results.  Nodes and weights are
chosen using known information about the shape of the integrand.  However,
classical quadrature methods have limited practical use for approximate Bayesian
inference.  Classical quadrature methods integrate over all unknown
parameters---not just hyperparameters---and the size of the integration grids
suffer from the curse of dimensionality, growing exponentially as parameters are
added to models.

More recent quadrature literature formalized theory and methods that yield
sparse integration grids, thereby mitigating the curse of dimensionality for
quadrature approximations of high dimensional integrals
\citep{Novak1996,Novak1999,Gerstner1998}. In statistics, sparse grid quadrature
methods have been used to approximate likelihoods that involve high dimensional
integrals, as can arise from econometric models \citep{Heiss2008}.  Sparse grid
quadrature has also been used to approximate posterior expectations, densities,
and integration constants for non-linear inverse problems with normal errors
\citep{Schillings2013, Emery2012}, estimate Kullback-Leibler information gains
to solve Bayesian experimental design problems \citep{Long2013}, and to
accelerate computations for specific non-linear Kalman filters
\citep{Arasaratnam2009,Jia2012}.  By comparison, we consider approximate
Bayesian posterior inference more generally.

We propose reformulating Bayesian posterior quantities, such as densities and
expectations, so that they can be efficiently approximated by combining
conditioning techniques with sparse grid quadrature methods.  Our reformulation
lets us apply sparse grid quadrature methods to hierarchical Bayesian models
with non-Gaussian structures and potentially many hyperparameters.  The
resulting computational approach greatly reduces computation time as compared
to MCMC approaches for many models, including fully non-Gaussian models.  Our
framework is very flexible and can be applied to other contexts.  For example,
it can also potentially be combined with INLA to allow fast inference
for latent Gaussian models with many hyperparameters.

We briefly review quadrature and sparse grid methods (\Cref{sec:sparse_grids}),
then introduce the Bayesian Inference via Sparse grid Quadrature Evaluation
(BISQuE) strategy to yield approximate inference for hierarchical Bayesian
models (\Cref{sec:wtdmix}).  Our method reduces the computational effort
required to approximate posterior densities, means, and variances in examples
where traditional MCMC methods are relatively slow (\Cref{sec:examples}).  We
conclude with discussions of extensions and other directions for future work
(\Cref{sec:discussion}).

\section{Quadrature and Sparse grid methods}
\label{sec:sparse_grids}

Let $f\paren\vx$ be a map from a $d$-dimensional space $\DS$ onto the real line
$\mathbb R$, and $w\paren\vx$ be a weight function with the same support.  The
integral
\begin{align}
  \label{eq:base_integral}
  I\paren f = \int_{\DS} f\paren\vx w\paren\vx d\vx
\end{align}
may be approximated via the weighted sum
\begin{align}
  \label{eq:base_quad}
  \hat I\paren f = \sum_{\ell=1}^{k_i} f\paren{\vx^{(i,\ell)}} w^{(i,\ell)}
\end{align}
for some choice of summation length $k_i\in\mathbb N$, nodes
$\mathcal A^i = \brc{\vx^{(i,\ell)} : \ell=1,\dots,k_i} \subset \DS$,
and weights
$\mathcal W^i = \brc{w^{(i,\ell)} : \ell=1,\dots,k_i}
  \subset \mathbb R^{k_i}$.
We will use the index $i$ shortly.
The approximation \autoref{eq:base_quad} is called a quadrature rule if the
integration domain $\DS$, weight function $w$, and desired approximation
accuracy or computational cost are used with specific procedures to
specify $k_i$, $\mathcal A^i$, and $\mathcal W^i$
\citep[Section 5.3]{Givens2013a}.
The number of nodes and weights $k_i$ balances
the approximation error in \autoref{eq:base_quad} with the approximation's
computational cost.  Large $k_i$ can yield more accurate approximation (or even
exact evaluation) of \autoref{eq:base_integral}, but at potentially high
computational cost. In practice, sequences of increasingly accurate quadrature
rules defined by
$\paren{k_1, \mathcal A^1, \mathcal W^1}$,
  $\paren{k_2, \mathcal A^2, \mathcal W^2}$, $\dots$
such that $k_1<k_2<\dots$ can be used to estimate and control approximation
error \citep{Laurie1985}. Quadrature rules can yield highly accurate
approximations for integrals $I\paren f$ of smooth functions $f$ defined on
$\DS$, but computational efficiency is difficult to achieve if $\DS$ has high
dimension.

For multidimensional $\DS$, product rules are the simplest quadrature rules to
construct, but these suffer from the curse of dimensionality.  Product
rules are formed by iteratively applying univariate quadrature rules along
each dimension of $\DS$ to approximate \autoref{eq:base_integral}; they are
aptly named because their nodes $\mathcal A^i $ are a Cartesian product of
nodes from the underlying univariate quadrature rules \citep[cf.][]{Novak1996}.
To be
precise, let $\DS$ be the product space $\DS = \DS_1\times\cdots\DS_d$ of
one-dimensional, $\sigma$-finite measure spaces $\DS_1,\dots,\DS_d$, and let
the weight function $w\paren\vx$ be the product
$w\paren\vx = \prod_{i=1}^d w_i\paren{x_i}$ of weight functions
$w_1\paren{x_1},\dots,w_d\paren{x_d}$ that are respectively defined on
$\DS_1,\dots,\DS_d$. If the target integral \autoref{eq:base_integral} is well
defined, then Fubini's theorem implies \autoref{eq:base_integral} may be
evaluated as an iterated integral.  Iterated integration allows approximation
by applying univariate quadrature rules along each dimension of
$\DS$.  Define $U_1^{i_1},\dots,U_d^{i_d}$ to be univariate
quadrature rules that respectively approximate integrals on $\DS_1,\dots,\DS_d$
with $k_{i_1},\dots,k_{i_d}$ nodes $\mathcal A_1^{i_1},\dots,\mathcal A_d^{i_d}$
and weights $\mathcal W_1^{i_1},\dots,\mathcal W_d^{i_d}$.  The product rule
that approximates \autoref{eq:base_integral} is defined via
\begin{align}
\label{eq:prod_rule}
  \paren{U_1^{i_1}\otimes\cdots\otimes U_d^{i_d}} \paren f =
  \sum_{\ell_1 = 1}^{k_{i_1}} \dots \sum_{\ell_d = 1}^{k_{i_d}}
  f\paren{x_1^{\paren{i_1,\ell_1}}, \dots, x_d^{\paren{i_d,\ell_d}}}
  w_1^{\paren{i_1,\ell_1}} \dots w_d^{\paren{i_d,\ell_d}}.
\end{align}
The product rule \autoref{eq:prod_rule} is a special case of the
general approximation form \autoref{eq:base_quad} because the nested sum in
\autoref{eq:prod_rule} may be re-expressed as a single sum over an enumeration
of the quadrature nodes
$({x_1^{\paren{i_1,\ell_1}}, \dots, x_d^{\paren{i_d,\ell_d}}}) \in
  \mathcal A_1^{i_1} \times \cdots \times \mathcal A_d^{i_d} $
that uses aggregated weights
$w_1^{\paren{i_1,\ell_1}} \dots w_d^{\paren{i_d,\ell_d}}$.
The product rule \autoref{eq:prod_rule} requires evaluation of $f$ at
$\abs{\mathcal A_1^{i_1} \times \cdots \times \mathcal A_d^{i_d}} =
  k_{i_1}\cdots k_{i_d}$
nodes.  The number of quadrature nodes grows exponentially as
$d\uparrow\infty$ if $f$ is explored equally in all dimensions, i.e.,
if $k_{i_1} = \cdots = k_{i_d}$.  The curse of dimensionality for product rules
can be partially mitigated by exploring $f$ unequally in different dimensions,
but this approach is only practical if $f$ is extremely smooth in some
dimensions.

By comparison, sparse grid quadrature rules are computationally efficient
approximations for integrals on multidimensional $\DS$.
\citet{Novak1996, Novak1999} use the \citet{Smolyak1963} formula to
combine univariate quadrature rules
$U_1^{i_1},\dots,U_d^{i_d}$ in a computationally efficient approximation
\autoref{eq:base_quad} of \autoref{eq:base_integral}.  The Smolyak formula
specifies a linear combination $A\paren{q,d}$ of product rules
\autoref{eq:prod_rule} that approximates \autoref{eq:base_integral} via
\begin{align}
\label{eq:smolForm}
  A\paren{q,d}\paren{f} = \sum_{q-d+1 \leq \abs{\v i} \leq q}
    \paren{-1}^{q-\abs{\v i}}
    \binom{d-1}{q - \abs{\v i}}
    \paren{U^{i_1}_1 \otimes \dots \otimes U^{i_d}_d}\paren f,
\end{align}
in which $q\geq d$ and $\abs{\v i}=i_1+\dots+i_d$.  The Smolyak rule
\autoref{eq:smolForm} is also a special case of the general approximation form
\autoref{eq:base_quad} because, as with \autoref{eq:prod_rule}, the underlying
quadrature nodes and weights may be enumerated and aggregated.
The constant $q\in\mathbb N$ is called the rule's
{\it level} and most directly controls the accuracy and computational cost of
the approximation in applications.  The Smolyak rule \autoref{eq:smolForm} is
called a sparse grid quadrature rule if each of the $j=1,\dots,d$ univariate
quadrature rules have nested nodes in the sense that
$\mathcal A_j^{1} \subset \mathcal A_j^{2} \subset \cdots$.
The rule \autoref{eq:smolForm} requires evaluation of $f$ at the nodes
\begin{align}
\label{eq:smol_nodes}
  \mathcal A\paren{q,d} = \bigcup_{q-d+1 \leq \abs{\v i} \leq q}
    \mathcal A_1^{i_1} \times \cdots \times A_d^{i_d}.
\end{align}
Adopting the convention that $A_j^0 = x_j^0$ for some base point
$x_j^0\in\DS_j$, nesting implies $\mathcal A\paren{q,d}$ is a sparse
subset of the nodes used by the product rule
$\paren{U^{q}_1 \otimes \dots \otimes U^{q}_d}\paren f$.

The sparse grid quadrature rule \autoref{eq:smolForm} mitigates the curse of
dimensionality by creating sparse integration grids relative to product rules,
but requires $f$ to satisfy stricter smoothness properties in exchange.
\citet{Novak1999} present growth rates, bounds, and approximations for the
number of quadrature nodes $k=\abs{\mathcal A\paren{q,d}}$ under different
scenarios. \citet{Novak1996} also show that the approximation error's order of
convergence is
\begin{align}
  \abs{I\paren f - A\paren{q,d}\paren f} = \mathcal O \paren{
    k^{-r} \paren{\log k}^{\paren{d-1}\paren{r/d+1}}
  }
\end{align}
if $f$ has a bounded mixed derivative $f^\paren{r,\dots,r}$.  Even more
precisely, \citet{Novak1999} show that $I\paren f = A\paren{q,d}\paren{f}$ if
$f$ is a polynomial with bounded total degree, i.e., that the approximation
\autoref{eq:smolForm} is {\it exact} for the integral
\autoref{eq:base_integral}.
In practice, the sparse grid quadrature rule \autoref{eq:smolForm} is most
computationally efficient for functions $f$ that behave approximately as
polynomials with relatively low total degree.  In statistical contexts, this is
similar to saying that
the rule \autoref{eq:smolForm} is most useful for polynomial surfaces $f$ that
are mainly driven by main effects and low order interaction terms.  We will
satisfy this requirement for computational efficiency in our application by
appealing, in part, to the Bayesian central limit theorem to claim that many
posterior surfaces and other quantities can be well approximated by the product
of a Gaussian weight function $w\paren\vx$ with a relatively low-order
correction term $f$.

\section{Posterior inference via weighted mixtures}
\label{sec:wtdmix}

We combine conditioning techniques with sparse grid quadrature rules to develop
specialized, computationally efficient formulas like \autoref{eq:smolForm} that
approximate Bayesian posterior inference for marginal quantities.  For example,
when used to approximate marginal posterior densities, our method will yield a
weighted mixture of full conditional posterior distributions.  Below, we briefly
motivate the Bayesian Inference via Sparse grid Quadrature Evaluation (BISQuE)
approximation strategy by arguing that it can be computationally inefficient to
use sparse grid quadrature rules to directly approximate posterior quantities
(\Cref{sec:bisque_motivation}).
First, our motivation simultaneously highlights the general strategy used to
apply sparse grid quadrature rules to Bayesian models as well as key technical
issues addressed by BISQuE.  Then, the remainder of \Cref{sec:wtdmix} defines
the family of posterior quantities to which BISQuE applies (\Cref{sec:model}),
the BISQuE approximation (\Cref{sec:inference}), and a nested integration
technique that is useful for applying BISQuE to models that lack closed form
expressions of posterior densities (\Cref{sec:factored_posteriors}).

\subsection{Motivation for BISQuE}
\label{sec:bisque_motivation}

Consider a generic hierarchical Bayesian model.
Let $\v X\in\Omega_0$ be a sample of continuous, discrete, or mixed
random variables from an arbitrary process.
Define a conditional probability model for $\v X$ such that
\begin{align}
\label{eq:basemodel}
\begin{split}
    \given{\v X}{\v\theta_1, \v\theta_2} \thicksim&~
        f\paren{\given{\v X}{\v\theta_1, \v\theta_2}} \\
    \paren{\v\theta_1, \v\theta_2} \thicksim&~
        f\paren{\v\theta_1, \v\theta_2}
\end{split}
\end{align}
for parameters $\v\theta_1\in\Omega_1$ and $\v\theta_2\in\Omega_2$.
Many Bayesian models can be written like \autoref{eq:basemodel}.
For example, many hierarchical Bayesian models add conditional independence
assumptions and hierarchical structure to \autoref{eq:basemodel} so that
\begin{align}
  f\paren{\given{\v X}{\v\theta_1, \v\theta_2}} =&
    f\paren{\given{\v X}{\v\theta_1}} \\
  f\paren{\v\theta_1, \v\theta_2} =&
      f\paren{\given{\v\theta_1}{\v\theta_2}}
      f\paren{\v\theta_2}.
\end{align}
Non-hierarchical models also fit within our framework
\autoref{eq:basemodel}.  For example, Bayesian formulations of some linear
regression models specify prior independence between regression coefficients
$\v\theta_1$ and variance components $\v\theta_2$, thus define
$f\paren{\v\theta_1, \v\theta_2} = f\paren{\v\theta_1} f\paren{\v\theta_2}$.

The marginal posterior density $f\paren{\given{\v\theta_1}{\v X}}$ is often
of interest in posterior inference.  The density may be computed by
integrating $\v\theta_2$ out of the joint posterior density
\begin{align}
\label{eq:direct_marginalizing}
  f\paren{\given{\v\theta_1}{\v X}} = \int
    f\paren{\given{\v\theta_1, \v\theta_2}{\v X}}
    d\v\theta_2.
\end{align}
Sparse grid quadrature rules \autoref{eq:smolForm} yield weighted-sum
approximations \autoref{eq:base_quad} of \autoref{eq:direct_marginalizing} by
introducing a weight function $w\paren{\v\theta_1,\v\theta_2,\v X}$ and
proceeding via
\begin{align}
\label{eq:direct_sparse}
  f\paren{\given{\v\theta_1}{\v X}} = \int
    \frac{ f\paren{\given{\v\theta_1, \v\theta_2}{\v X}} }
         {w\paren{\v\theta_1,\v\theta_2,\v X}}
    w\paren{\v\theta_1,\v\theta_2,\v X}
    d\v\theta_2
  \approx
  \sum_{\ell=1}^{k_i}
  \frac{ f\paren{\given{\v\theta_1, \v\theta_2^{\paren{i,\ell}}}{\v X}} }
       {w\paren{\v\theta_1, \v\theta_2^{\paren{i,\ell}},\v X}}
  w^{\paren{i,\ell,\v\theta_1}},
\end{align}
in which quadrature nodes $\v\theta_2^{\paren{i,\ell}}$ and weights
$w^{\paren{i,\ell,\v\theta_1}}$ are determined by applying the Smolyak formula
\autoref{eq:smolForm} to a collection of univariate quadrature rules that are
appropriate for the support of $\v\theta_2$.
For fixed $\v\theta_1\in\Omega_1$, the Gaussian approximation to
$f\paren{\given{\v\theta_1, \v\theta_2}{\v X}}$ will often be a sensible
default choice for the weight function $w\paren{\v\theta_1,\v\theta_2,\v X}$
since the weight ratio $f/w$ in \autoref{eq:direct_sparse} accounts for
deviations from normality in $f\paren{\given{\v\theta_1, \v\theta_2}{\v X}}$.

The direct marginal posterior density approximation \autoref{eq:direct_sparse}
has two key inefficiencies that the BISQuE approximation completely avoids or
minimizes. First, the weight function $w$ depends on $\v\theta_1$, which
implies a separate weight function must be used to approximate
$f\paren{\given{\v\theta_1}{\v X}}$ at each $\v\theta_1\in\Omega_1$.  Second,
the approximation \autoref{eq:direct_sparse} assumes
$f\paren{\given{\v\theta_1, \v\theta_2}{\v X}}$ is computable.  Oftentimes,
the joint posterior density $f\paren{\given{\v\theta_1, \v\theta_2}{\v X}}$ is
only known in closed form up to a proportionality constant because the
density's integration constant requires numerical approximation for many
Bayesian models.  While sparse grid quadrature rules could approximate the
integration constant, BISQuE is able to avoid or reduce computational cost of
the approximation.

\subsection{Posterior quantities targeted by BISQuE}
\label{sec:model}

We develop BISQuE to approximate marginal posterior quantities
$h\paren{\v\theta_1;\v X}$ of hierarchical models \autoref{eq:basemodel} that
are defined implicitly with respect to a function or random variable
$h\paren{\v\theta_1, \v\theta_2; \v X}$ via
\begin{align}
\label{eq:post_qty}
  h\paren{\v\theta_1;\v X} =& \int
    h\paren{\v\theta_1, \v\theta_2;\v X}
    f\paren{\given{\v\theta_2}{\v X}} d \v\theta_2.
\end{align}
For example, the construction \autoref{eq:post_qty} defines the marginal
posterior density $h\paren{\v\theta_1;\v X} = f\paren{\given{\v\theta_1}{\v X}}$
when
$h\paren{\v\theta_1, \v\theta_2; \v X} =
  f\paren{\given{\v\theta_1}{\v\theta_2,\v X}}$.
The posterior marginal density $f\paren{\given{\v\theta_2}{\v X}}$
and all other marginal posterior quantities may be formed by switching the
roles of $\v\theta_1$ and $\v\theta_2$.
In comparison to the definition \autoref{eq:direct_marginalizing} used in the
direct sparse grid approximation \autoref{eq:direct_sparse}, the BISQuE
construction \autoref{eq:post_qty} uses conditioning to express the joint
posterior density in conditional form, as
$f\paren{\given{\v\theta_1,\v\theta_2}{\v X}} =
  f\paren{\given{\v\theta_1}{\v\theta_2,\v X}}
  f\paren{\given{\v\theta_2}{\v X}}$.
The construction \autoref{eq:post_qty}
allows us to develop sparse grid quadrature rules with weight functions
$w\paren{\v\theta_2,\v X}$ that only depend on $\v\theta_2$
(\Cref{sec:inference}), thus addresses the first technical issue described
in \Cref{sec:bisque_motivation}.

The BISQuE construction \autoref{eq:post_qty} allows one set of
quadrature nodes and weights to be reused to approximate many posterior
quantities.  For example, \autoref{eq:post_qty} defines the posterior mean
$h\paren{\v\theta_1; \v X} = E\brk{\given{g\paren{\v\theta_1}}{\v X}}$
when
$h\paren{\v\theta_1, \v\theta_2; \v X} =
  \text{E}\brk{\given{g\paren{\v\theta_1}}{\v\theta_2, \v X}}$.
Again, the approach relies on conditioning as
\begin{align}
\label{eq:post_rv}
\begin{split}
  \text{E}\brk{\given{g\paren{\v\theta_1}}{\v X}} =&
    \text{E}_{\given{\v\theta_2}{\v X}}
    \brc{\text{E}\brk{\given{g\paren{\v\theta_1}}{\v\theta_2, \v X}}} \\
  =& \int \text{E}\brk{\given{g\paren{\v\theta_1}}{\v\theta_2, \v X}}
    f\paren{\given{\v\theta_2}{\v X}}
    d \v\theta_2.
\end{split}
\end{align}
Posterior predictive distributions, variances and higher central moments,
cumulative distribution functions, and model selection criteria such as the
deviance information criteria \citep[DIC,][]{Spiegelhalter2002} and the
Watanabe-Akaike information criterion \citep[WAIC,][]{Watanabe2010} can also
be expressed through one or more applications of \autoref{eq:post_qty}.
For example, the posterior variance
$\text{Var}\paren{\given{g\paren{\v\theta_1}}{\v X}}$
can be approximated by using the law of total variance to introduce
expectations with respect to $f\paren{\given{\v\theta_2}{\v X}}$ via
\begin{align}
\label{eq:total_variance}
  \text{Var}\paren{\given{g\paren{\v\theta_1}}{\v X}} =&
    \text{E}_{\given{\v\theta_2}{\v X}}\brk{
      \text{Var}\paren{\given{g\paren{\v\theta_1}}{\v\theta_2, \v X}}
    } + \\ &
    \text{E}_{\given{\v\theta_2}{\v X}}\brk{\paren{
      \text{E}\brk{\given{g\paren{\v\theta_1}}{\v\theta_2, \v X}} -
      \text{E}\brk{\given{g\paren{\v\theta_1}}{\v X}}}^2
    },
\end{align}
for which
\begin{align}
\label{eq:total_variance_h}
  h\paren{\v\theta_1, \v\theta_2; \v X} =
    \text{Var}\paren{\given{g\paren{\v\theta_1}}{\v\theta_2, \v X}} +
    \paren{
      \text{E}\brk{\given{g\paren{\v\theta_1}}{\v\theta_2, \v X}} -
      \text{E}\brk{\given{g\paren{\v\theta_1}}{\v X}}}^2.
\end{align}
Note that here the marginal posterior expectation
$\text{E}\brk{\given{g\paren{\v\theta_1}}{\v X}}$ must be approximated before
\autoref{eq:total_variance}.  We present expressions for the other
quantities in Supplement \nolink{\Cref{supp:sec:post_qty_examples}}.

\subsection{Approximate posterior inference via BISQuE}
\label{sec:inference}

We modify the integral form \autoref{eq:base_integral} to enable the use of
sparse grid quadrature rules \autoref{eq:smolForm} to approximate
marginal posterior quantities \autoref{eq:post_qty} of hierarchical Bayesian
models \autoref{eq:basemodel}.  While we define marginal posterior quantities
by integrating functions over the posterior density
$f\paren{\given{\v\theta_2}{\v X}}$, numerical
integration methods often use transformations to increase computational
stability and efficiency.  Thus, we develop quadrature rules that integrate over
$f\paren{\given{\v\nu}{\v X}}$ where $\v\nu = T\paren{\v\theta_2}\in\mathbb R^p$
is defined by a monotone transformation to a real coordinate space
$T:\Omega_2\rightarrow\mathbb R^p$.  Consider a transformed density
\begin{align}
  \label{eq:tx_dens}
  f\paren{\given{\v\nu}{\v X}} =
    f\paren{\given{\inv T\paren{\v\nu}}{\v X}}
    \abs{J\paren{\inv{T}\paren{\v\nu}}},
\end{align}
where $\abs{J\paren{\inv T\paren{\v\nu}}}$ is the determinant of the
Jacobian for the transformation $\inv T$.  We propose using sparse grid
quadrature rules \autoref{eq:smolForm} to derive quadrature nodes and weights
that approximate marginal posterior quantities \autoref{eq:post_qty} via
the BISQuE approximation
\begin{align}
\label{eq:baseapprox}
  h\paren{\v\theta_1; \v X} =&
      \int
        h\paren{\v\theta_1, \inv T\paren{\v\nu}; \v X}
        \frac{f\paren{\given{\v\nu}{\v X}}}
             {w\paren{\v\nu, \v X}}
        w\paren{\v\nu, \v X}
        d \v\nu \\
    \approx&
    \sum_{\ell=1}^{k_i}
      h\paren{\v\theta_1, \v\theta_2^{(i,\ell)}; \v X}
      \tilde w^{(i,\ell)},
\end{align}
in which
\begin{align}
  \label{eq:wtdef}
    \tilde w^{(i,\ell)} =
        \frac{f\paren{\given{\v\nu^{(i,\ell)}}{\v X}}}
            {w\paren{\v\nu^{(i,\ell)},\v X}}
        w^{(i,\ell)},
\end{align}
$w\paren{\v\nu,\v X}$ is a weight function; and $w^{(i,\ell)}$,
$\v\nu^{(i,\ell)}$, and $\v\theta_2^{(i,\ell)} = \inv T\paren{\v\nu^{(i,\ell)}}$
are respectively quadrature weights, nodes, and back-transformed nodes.

Sparse grid quadrature theory implies the computational efficiency of the
approximation \autoref{eq:baseapprox} relies on several statistical and
numerical assumptions.  The weight function $w\paren{\v\nu,\v X}$ should
approximate the transformed density $f\paren{\given{\v\nu}{\v X}}$ and
have known, computationally efficient, nested quadrature rules.  In particular,
such quadrature rules have been developed for Gaussian weight functions
\citep{Genz1996}.  Thus, we appeal to Bayesian analogs of the central limit
theorem to justify proposing the Gaussian approximation
$f^G\paren{\given{\v\nu}{\v X}}$ at the posterior mode of
$f\paren{\given{\v\nu}{\v X}}$ as a sensible default choice for a weight
function for many Bayesian models.  This approximation holds if the same size
is large, the dimension of the model is fixed, and both the prior and likelihood
are twice differentiable near the mode of the posterior distribution
\citep[pg. 224--225]{Berger1985}.  Sparse grid quadrature rules will also be
most efficient if the integrand
$h\paren{\v\theta_1, \inv T\paren{\v\nu}; \v X}
 f\paren{\given{\v\nu}{\v X}} /
 w\paren{\v\nu, \v X}$
in \autoref{eq:baseapprox} can be well-approximated by a low-order polynomial
in $\v\nu$.  This requirement is easier to satisfy if $w\paren{\v\nu,\v X}$
approximates $f\paren{\given{\v\nu}{\v X}}$ well and
$h\paren{\v\theta_1, \inv T\paren{\v\nu}; \v X}$ is slowly varying with respect
to $\v\nu$.

Standardizing the BISQuE approximation \autoref{eq:baseapprox} weights
$\tilde w^{(i,\ell)}$ can address part of the second technical issue described
in \Cref{sec:bisque_motivation}.  For example, in some
Bayesian models both the joint $f\paren{\given{\v\theta_1,\v\theta_2}{\v X}}$
and marginal $f\paren{\given{\v\theta_2}{\v X}}$ posterior densities are known
only up to a proportionality constant, but the full conditional posterior
$f\paren{\given{\v\theta_1}{\v\theta_2, \v X}}$ is available in closed form
(\Cref{sec:examples}).  Marginal posterior probabilities and expectations
cannot be computed without either approximating the proportionality constant or
using numerical approximation techniques that implicitly cancel the constant.
We propose using standardized weights
$\tilde w_*^{(i,\ell)} =
  \tilde w^{(i,\ell)} / \sum_{j=1}^{k_i} \tilde w^{(i,j)}$
that sum to one in order to approximate marginal posterior quantities
$h\paren{\v\theta_1;\v X}$ like $f\paren{\given{\v\theta_1}{\v X}}$ by
implicitly cancelling the unknown integration constants.  The result borrows
ideas from importance sampling \citep[pg. 181]{Givens2013a}.
An alternate definition for posterior quantities,
\begin{align}
  \label{eq:stdapprox}
  h\paren{\v\theta_1; \v X} =&
    \frac{
      \int
        h\paren{\v\theta_1, \v\theta_2;\v X}
        f\paren{\given{\v\theta_2}{\v X}} d \v\theta_2
    }{
      \int f\paren{\given{\v\theta_2}{\v X}} d \v\theta_2
    },
\end{align}
is equivalent to the original construction \autoref{eq:post_qty} since
$\int f\paren{\given{\v\theta_2}{\v X}} d \v\theta_2 = 1$.
Plugin BISQuE approximations \autoref{eq:baseapprox} for the numerator and
denominator in \autoref{eq:stdapprox} yield quadrature approximations with
standardized weights via
\begin{align}
\label{eq:std_wtd_approx}
  \frac{
    \int
      h\paren{\v\theta_1, \v\theta_2;\v X}
      f\paren{\given{\v\theta_2}{\v X}} d \v\theta_2
  }{
    \int f\paren{\given{\v\theta_2}{\v X}} d \v\theta_2
  }
  \approx
    \frac{
      \sum_{\ell=1}^{k_i} h\paren{\v\theta_1, \v\theta_2^{(i,\ell)}; \v X}
        \tilde w^{\paren{i,\ell}}
    }{
      \sum_{j=1}^{k_i} \tilde w^{\paren{i,j}}
    }
  =
    \sum_{\ell=1}^{k_i} h\paren{\v\theta_1, \v\theta_2^{(i,\ell)}; \v X}
      \tilde w_*^{(i,\ell)}.
\end{align}
Standardization also allows approximations of
$f\paren{\given{\v\theta_1}{\v X}}$ to integrate exactly to one.

\Cref{table:outline} summarizes the BISQuE approach outlined in this section as
it would be applied when using a Gaussian approximation to the transformed
posterior density to approximate posterior quantities \autoref{eq:post_qty}.

\begin{table}
  \centering
  \caption{Summary of steps to develop a BISQuE approximation.}
  \label{table:outline}
  \ra{1.5}
  \begin{tabularx}{\linewidth}{cX}
    \toprule
    1. & Write posterior quantity of interest in BISQuE form
      \autoref{eq:post_qty}. \vspace{.5em} \\

    & \hspace{.5em}\begin{minipage}{.95\linewidth}
    {\it Computable approximations or exact expressions
          must exist for the components $h\paren{\v\theta_1, \v\theta_2; \v X}$ and
          $f\paren{\given{\v\theta_2}{\v X}}$. \Cref{sec:factored_posteriors}
          proposes nested integration strategies \autoref{eq:factored_t2} and
          \autoref{eq:factored_t1} if approximation is necessary; nested
          Laplace approximations can also be used for components in latent Gaussian
          models \citep[cf.][]{Rue2009a}.}
      \end{minipage} \vspace{1em} \\

    2. & Select transformation $\v\nu = T\paren{\v\theta_2}$ to map
      $\v\theta_2\in\Omega_2$ to $\v\nu\in\mathbb R^p$. \vspace{.5em} \\

    & \hspace{.5em}\begin{minipage}{.95\linewidth}
    {\it  Favor transformations $T$ that yield an approximately Gaussian
    posterior density $f\paren{\given{\v\nu}{\v X}}$.}
      \end{minipage} \vspace{1em} \\

    3. & Apply the BISQuE approximation that uses unstandardized
      \autoref{eq:baseapprox} or standardized \autoref{eq:std_wtd_approx}
      weights. \vspace{.5em} \\

    & \hspace{.5em}\begin{minipage}{.95\linewidth}
    {\it The level $q\in\mathbb N$ of the underlying sparse grid quadrature rule
    \autoref{eq:smolForm} determines the integration nodes $\v\nu^{(i,\ell)}$
    and weights $w^{(i,\ell)}$.}
    \end{minipage} \vspace{1em} \\

    4. & Increase the level $q$ of underlying quadrature rule
      \autoref{eq:smolForm} until the approximation \autoref{eq:baseapprox} or
      \autoref{eq:std_wtd_approx} converges. \vspace{.5em} \\

    & \hspace{.5em}\begin{minipage}{.95\linewidth}
    {\it Nested quadrature rules allow the level $q$ approximation to reduce
    computational cost by reusing quadrature nodes and weight ratios from the
    level $q-1$ approximation.}
    \end{minipage} \vspace{.5em} \\

    \bottomrule
  \end{tabularx}
\end{table}

\subsection{Nested integration strategies for BISQuE}
\label{sec:factored_posteriors}

While hierarchical Bayesian models \autoref{eq:basemodel} typically have
closed form expressions for the likelihood
$f\paren{\given{\v X}{\v\theta_1,\v\theta_2}}$ and prior
$f\paren{\v\theta_1,\v\theta_2}$, many models do not have closed form
expressions for the posterior densities $f\paren{\given{\v\theta_2}{\v X}}$ and
$f\paren{\given{\v\theta_1}{\v\theta_2, \v X}}$.  Lack of closed form
expressions is a concern related to the second technical issue described
in \Cref{sec:bisque_motivation}.  We propose a nested numerical integration
scheme to address the concern and allow application of BISQuE to a wider range
of models.  Recall that for a fixed dataset $\v X$, the joint posterior density
$f\paren{\given{\v\theta_1,\v\theta_2}{\v X}}$ is often only known up to a
proportionality constant since
\begin{align}
  f\paren{\given{\v\theta_1,\v\theta_2}{\v X}} =
    \frac{f\paren{{\v\theta_1,\v\theta_2,\v X}}}{f\paren{\v X}}
  \propto
    f\paren{\given{\v X}{\v\theta_1,\v\theta_2}}
    f\paren{{\v\theta_1,\v\theta_2}}
\end{align}
and the marginal density $f\paren{\v X}$ often requires prohibitively expensive
numerical approximation.

The densities $f\paren{\given{\v\theta_2}{\v X}}$ and
$f\paren{\given{\v\theta_1}{\v\theta_2, \v X}}$ may be derived (and ultimately
approximated) indirectly, by factoring the joint density
$f\paren{\v\theta_1,\v\theta_2,\v X}$ into components
$g_1\paren{\v\theta_1, \v\theta_2; X}$ and $g_2\paren{\v\theta_2; \v X}$ such
that
\begin{align}
\label{eq:joint_decomp}
  f\paren{\v\theta_1,\v\theta_2,\v X} =
    g_1\paren{\v\theta_1, \v\theta_2; \v X}
    g_2\paren{\v\theta_2; \v X}.
\end{align}
The factored joint density \autoref{eq:joint_decomp} implies
\begin{align}
\label{eq:factored_t2}
  f\paren{\given{\v\theta_2}{\v X}} =&
    \int f\paren{\given{\v\theta_1,\v\theta_2}{\v X}} d\v\theta_1
  = \frac{g_2\paren{\v\theta_2; \v X} C_1\paren{\v\theta_2}}{f\paren{\v X}}
\end{align}
and
\begin{align}
\label{eq:factored_t1}
  f\paren{\given{\v\theta_1}{\v\theta_2, \v X}} =&
    \frac{f\paren{\given{\v\theta_1,\v\theta_2}{\v X}}}
         {f\paren{\given{\v\theta_2}{\v X}}}
  = \frac{g_1\paren{\v\theta_1,\v\theta_2;\v X}}
          {C_1\paren{\v\theta_2}},
\end{align}
for which the integration constant $C_1\paren{\v\theta_2}$ must be approximated
numerically and is specified via
\begin{align}
\label{eq:factored_int}
  C_1\paren{\v\theta_2} =
    \int g_1\paren{\v\theta_1,\v\theta_2;\v X} d\v\theta_1.
\end{align}

The alternate expressions \autoref{eq:factored_t2} and \autoref{eq:factored_t1}
allow BISQuE to approximate posterior inference for models that lack closed
form expressions for the densities $f\paren{\given{\v\theta_2}{\v X}}$ and
$f\paren{\given{\v\theta_1}{\v\theta_2, \v X}}$.  Standardized BISQuE weights
$\tilde w_*^{\paren{i,\ell}}$ implicitly cancel the unknown factor
$f\paren{\v X}$, and standard quadrature techniques
can efficiently approximate the integration constant \autoref{eq:factored_int}
when the parameter vector $\v\theta_1$ has small dimension.  The parameters
$\v\theta_1$ and $\v\theta_2$ can often be defined or repartitioned to satisfy
this requirement because the hierarchical model \autoref{eq:basemodel} places
few restrictions on the parameters; we use this flexibility in
\Cref{sec:examples}.  The added computational cost that the nested integration
\autoref{eq:factored_int} adds to the BISQuE approximation is minimized as the
integration constant \autoref{eq:factored_int} only needs to be approximated
relatively few times, specifically, at the quadrature nodes and when
developing the weight function---e.g., the Gaussian approximation at the
posterior mode.

\section{Examples}
\label{sec:examples}

We demonstrate the benefits of the BISQuE approximation \autoref{eq:baseapprox}
on data that are typically analyzed with standard, Gibbs sampling techniques
for approximate Bayesian posterior inference.  We approximate posterior
inference for a fully non-Gaussian capture-recapture model
(\Cref{sec:fur_seals_ex}), a spatial Gaussian process model
(\Cref{sec:spatial_main}), and a more complex, applied spatial Gaussian process
model for climate teleconnection (\Cref{sec:resp_ex}).  Posterior distributions
in the first and third examples respectively require integration over 8 and
5-dimensional parameter vectors $\v\theta_2$.  Posterior approximations for the
second and third examples have computational complexity that is
$\mathcal O\paren{MN^3}$ in the number of spatial observations $N$ and $M$
points at which the posterior distribution is explored, thus computational
strategies like BISQuE that reduce the number of points required for
posterior approximation can be extremely beneficial.

We compare posterior inference and computational effort between standard Gibbs
sampling techniques and BISQuE.  Computational effort is measured indirectly
with respect to computation time.  All computations are conducted on
a modest workstation with eight logical processors.  We use parallelization to
compute the $k_i$ mixture components of the BISQuE approximation and to draw
posterior predictive samples via composition sampling in the spatial examples
\citep[cf.][pg. 126]{Banerjee2015}.  For each posterior quantity, the level $q$
for the underlying sparse grid quadrature rule \autoref{eq:smolForm} is chosen
to be the smallest value (i.e., the simplest approximation) such that the
posterior density approximations have converged.  The number of Gibbs steps
used in each approximation is similarly chosen.  The BISQuE
approximation also requires specification of univariate quadrature rules, for
which we choose nested Gauss-Hermite rules \citep{Genz1996}.

\subsection{Fur seals}
\label{sec:fur_seals_ex}

\subsubsection{Data and model}

\citet[][example 7.7]{Givens2013a} analyze data from a capture-recapture study
conducted in New Zealand.  The study's research goal was to estimate the total
number of pups in a fur seal colony $N\in\mathbb N$.  Researchers visited the
colony $I=7$ times throughout the course of a single season.  In each visit, the
researchers captured and marked all of the fur seal pups present, noting the
total number of pups captured in each visit
$\v c = \paren{c_1,\dots,c_I} \in\mathbb N^I$ in addition to the number of
newly captured pups $m_1,\dots,m_I \in\mathbb N$.  The data are analyzed using
a Bayesian model for capture-recapture data \autoref{eq:pupmodel}, and
posterior distributions are approximated with a Gibbs sampler.  Gibbs sampling
is particularly inefficient for this model as one pair of hyperparameters has
high posterior correlation and are only weakly identified  by the data.  By
comparison, the BISQuE strategy \autoref{eq:baseapprox} approximates posterior
quantities for this model with substantially less computational effort.

The model \autoref{eq:pupmodel} assumes the total population size $N$ remains
fixed during the time
period of the study (i.e., the model assumes a closed population). Let
$r=\sum_{i=1}^I m_i$ be the total number of pups captured during the study.
\citet{Givens2013a} introduce a vector
$\v\alpha=\paren{\alpha_1,\dots,\alpha_I} \in \brk{0,1}^I$
with capture probabilities for each census attempt and discuss modeling the
data with the hierarchical model
\begin{align}
\label{eq:pupmodel}
\begin{split}
    f\paren{\given{\v c, r}{N, \v\alpha}} \propto&\,
        \frac{N!}{\paren{N-r}!}
        \prod_{i=1}^I \alpha_i^{c_i} \paren{1-\alpha_i}^{N-c_i} \\
    f\paren{N} \propto&\, 1/N \\
    f\paren{\given{\alpha_i}{\theta_1, \theta_2}} \thicksim&\,
        \text{Beta}\paren{\theta_1, \theta_2} \text{ for } i=1,\dots,I \\
    f\paren{\theta_1,\theta_2} \propto&
      \exp\brc{-\paren{\theta_1+\theta_2}/1000},
\end{split}
\end{align}
in which $\paren{\theta_1,\theta_2}$ are hyperparameters for the capture
probabilities.  We use the Beta distribution's mean--sample size
parameterization to increase the identifiability of the hyperparameters.
Specifically, let $U_1 = \logit \paren{\theta_1/\paren{\theta_1+\theta_2}}$
and $U_2 = \log \paren{\theta_1 + \theta_2}$ and fix $U_2=5.5$.

\subsubsection{Posterior inference and results}

\citet{Givens2013a} use standard Gibbs-sampling approaches to
draw posterior samples for model parameters.  The full conditional posterior
distributions $f\paren{\given{N}{\v c, r, \v\alpha, \theta_1, \theta_2}}$ and
$f\paren{\given{\v\alpha}{\v c, r, N, \theta_1, \theta_2}}$ are conjugate and
easy to sample.  Posterior samples for $U_1$ are drawn using Metropolis steps.
The sampler is run for 100,000 iterations, taking 298 seconds to complete;
posterior inference uses the final 50,000 samples.

We use the BISQuE strategy to approximate the posterior marginal densities
$f\paren{\given{N}{\v c, r}}$, $f\paren{\given{\alpha_i}{\v c, r}}$,
and $f\paren{\given{U_1}{\v c, r}}$.  \Cref{table:fur_seal_mapping} connects
this example's notation to that used with BISQuE.  When used as the BISQuE
conditioning variable $\v\theta_2$,  we map parameters to the real line by
using log transforms with $N-r$ and logit transforms with the capture
probabilities $\v\alpha$.  We also rely on the Gaussian approximation to the
negative binomial distribution in order to justify using $N$ as a conditioning
variable $\v\theta_2$ in BISQuE.  Almost all conditional and marginal posterior
densities required for BISQuE are computable in closed form up to a
proportionality constant; refer to \citet[eqs. 7.16, 7.17]{Givens2013a}
and Supplement \nolink{\Cref{supp:sec:fur_seals}} for details.  The posterior
for $f\paren{\given{U_1}{\v c, r}}$ requires approximation via nested
integration strategies (\Cref{sec:factored_posteriors}).

Posterior inference via BISQuE is effectively identical to posterior
inference via Gibbs sampling, but is computed with substantially less effort.
Gibbs sampling takes 298 seconds to complete on our test machine, whereas the
BISQuE approximations require a total of 5 seconds
(\Cref{table:fur_seal_mapping}), and posterior densities are nearly identical
(\Crefrange{fig:post_N}{fig:post_alphas_plot}).

\begin{table*}\centering
\caption{Definitions of the parameters and posterior quantities for the
  BISQuE approximations in \Cref{sec:examples}. $\v X = \paren{\v c ,r}$ for
  the fur seals example (\Cref{sec:fur_seals_ex}), and $\v X = \v Y$ for the
  Remote effects spatial process model example (RESP, \Cref{sec:resp_ex}). The
  marginal posterior densities for the covariance parameters
  $\paren{\sigma^2, \rho, \nu}$ in the spatial example (\Cref{sec:spatial_main})
  are computed using sparse-grid quadrature methods \autoref{eq:smolForm} to
  directly marginalize the joint posterior distribution
  $f\paren{\given{\sigma^2,\rho,\nu}{\v X}}$ at each evaluation point.
  Computation times are also presented. For the RESP example, let
  $\v\theta^* = \paren{\sigma^2_w, \sigma^2_\varepsilon, \sigma^2_\alpha,
    \rho_w, \rho_\alpha}$ and
  $I\paren\vs = (c_{i-1}\paren\vs, c_{i}\paren\vs)$. }
\label{table:fur_seal_mapping}
\ra{2}
\begin{tabular}{ccccccrr}

&&&&& \multicolumn{2}{c}{Time (sec.)} \\

Example &
$h\paren{\v\theta_1; \v X}$ &
$h\paren{\v\theta_1, \v\theta_2; \v X}$ &
$\v\theta_1$ &
$\v\theta_2$ &
BISQuE & Gibbs \\

\midrule

\multirow{3}{*}{Fur seals} &
$f\paren{\given{N}{\v c, r}}$ &
$f\paren{\given{N}{\v\theta_2, \v c, r}}$ &
$N$ &
$\paren{\v\alpha, U_1}$ & 0.3 & 298 \\

& $f\paren{\given{\alpha_i}{\v c, r}}$ &
$f\paren{\given{\alpha_i}{\v\theta_2, \v c, r}}$ &
$\alpha_i$ &
$\paren{N,U_1}$ & 0.1 & 298 \\

& $f\paren{\given{U_1}{\v c, r}}$ &
$f\paren{\given{U_1}{\v\theta_2, \v c, r}}$ &
$U_1$ &
$\v\alpha$ & 5.0 & 298 \\

\midrule

\multirow{6}{*}{Spatial} &
$\text{E}\brk{\given{\v X_0}{\v X}}$ &
$\text{E}\brk{\given{\v X_0}{\v\theta_2, \v X}}$ &
$\v X_0$ &
$\paren{\sigma^2, \rho, \nu}$ & 6 & 2,651 \\

&
$\text{Var}\paren{\given{\v X_0}{\v X}}$ &
\autoref{eq:total_variance_h} &
$\v X_0$ &
$\paren{\sigma^2, \rho, \nu}$ & 6 & 2,651 \\

&
$f\paren{\given{\v X_0}{\v X}}$ &
$f\paren{\given{\v X_0}{\v\theta_2, \v X}}$ &
$\v X_0$ &
$\paren{\sigma^2, \rho, \nu}$ & 6 & 2,651 \\

& $f\paren{\given{\sigma^2}{\v X}}$ &
N/A &
$\sigma^2$ &
$\paren{\rho, \nu}$ & 74 & 2,043 \\

& $f\paren{\given{\rho}{\v X}}$ &
N/A &
$\rho$ &
$\paren{\sigma^2, \nu}$ & 74 & 2,043 \\

& $f\paren{\given{\nu}{\v X}}$ &
N/A &
$\nu$ &
$\paren{\sigma^2, \rho}$ & 74 & 2,043 \\

\midrule

\multirow{2}{*}{RESP} &
$f\paren{\given{\v Y_0}{\v Y}}$ &
$f\paren{\given{\v Y_0}{\v\theta_2, \v Y}}$ &
$\v Y_0$ &
$\v\theta^*$ & 118 & 9,086  \\

& $f({{\tilde{\v Y}_0}\vert{\v Y}})$ &
$P( Y_0\paren{\vs,t} \in I\paren\vs  \vert
  \v\theta_2, \v Y)$ &
$\v Y_0$ &
$\v\theta^*$ & 118 & 9,086

\end{tabular}
\end{table*}

\begin{knitrout}
\definecolor{shadecolor}{rgb}{0.969, 0.969, 0.969}\color{fgcolor}\begin{figure}

{\centering \includegraphics[width=\maxwidth]{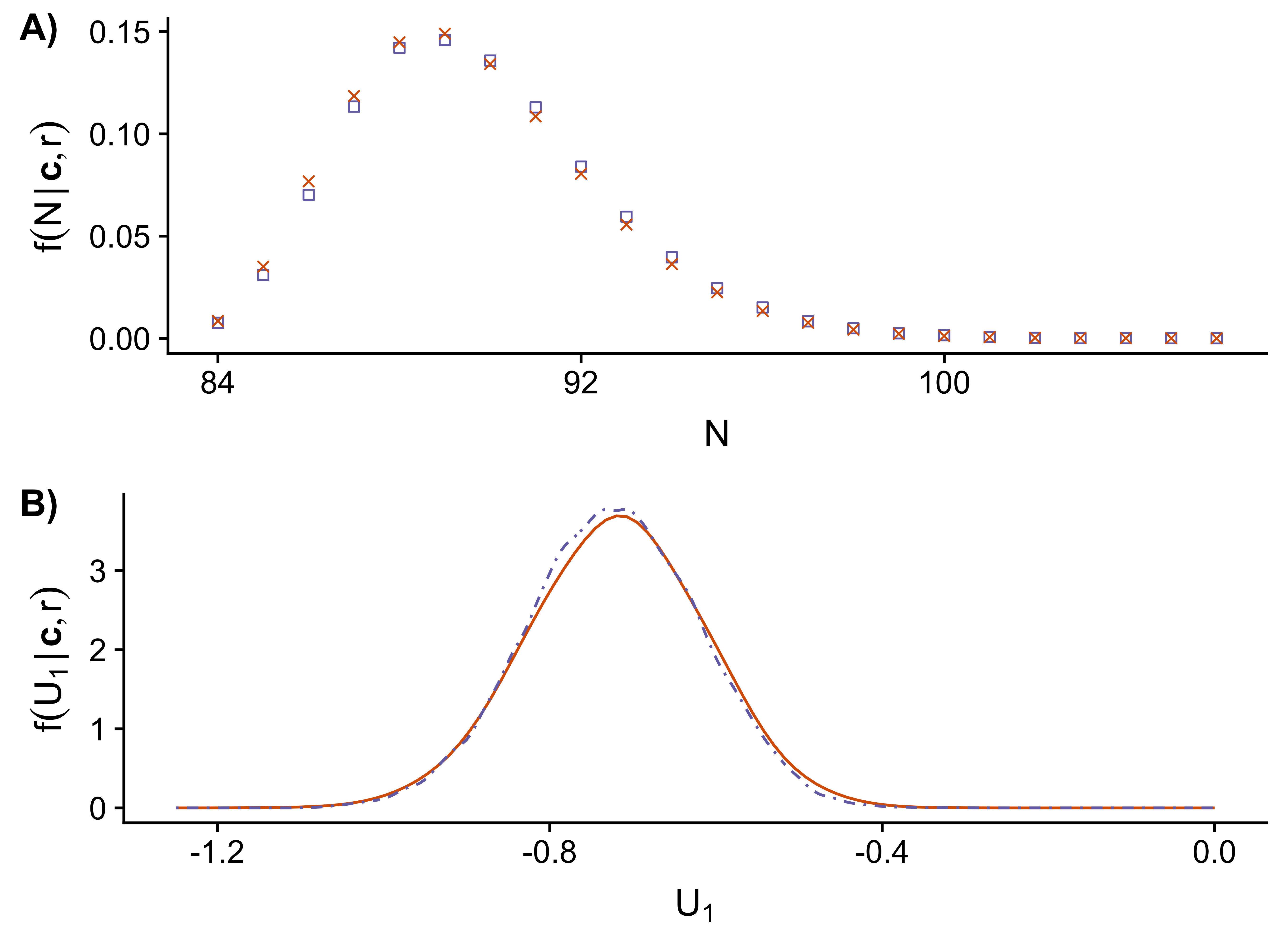} 

}

\caption{Fur seals example: A) BISQuE ({\color{dkrd}x}) and Gibbs ({\color{dkpu}$\square$}) approximations to the posterior density for total number of fur seal pups $f\paren{\given{N}{\v c, r}}$ are nearly identical. B) BISQuE ({\color{dkrd}---}) and Gibbs ({\color{dkpu}-$\cdot$-}) approximations to the joint posterior density $f\paren{\given{U_1}{\v c, r}}$ are nearly identitcal.}\label{fig:post_N}
\end{figure}

\end{knitrout}

\begin{knitrout}
\definecolor{shadecolor}{rgb}{0.969, 0.969, 0.969}\color{fgcolor}\begin{figure}

{\centering \includegraphics[width=\maxwidth]{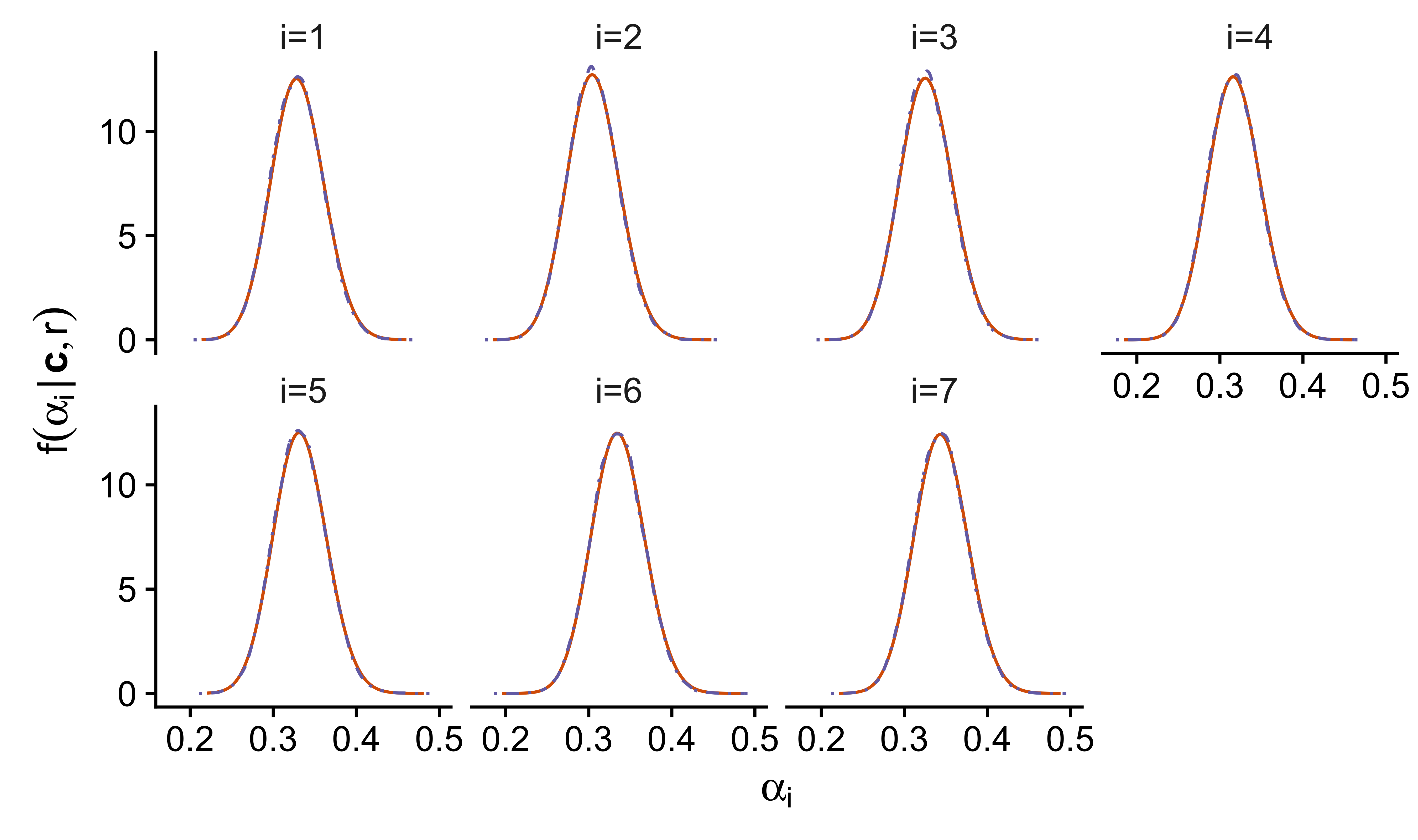} 

}

\caption[Fur seals example]{Fur seals example: BISQuE ({\color{dkrd}---}) and Gibbs ({\color{dkpu}-$\cdot$-}) approximations to the posterior densities $f\paren{\given{\alpha_i}{\v c, r}}$ are nearly identitcal.}\label{fig:post_alphas_plot}
\end{figure}

\end{knitrout}

\subsection{Spatial}
\label{sec:spatial_main}

\subsubsection{Simulated data and model}
\label{sec:spatial_ex}

We work with data simulated from a geostatistical spatial model.  Gibbs sampling
is computationally expensive for such models because it involves decomposing
spatially-structured covariance matrices in $\mathbb R^{N\times N}$ at each
Gibbs iteration, where $N$ is the number of observations.
Let $\brc{X\paren\vs}_{\vs\in\D}$ be a random field, whose stochasticity is
defined by a mean-zero Gaussian process on a
continuous spatial domain $\D\subset\mathbb R^2$.  Let the covariance
$\Cov{X\paren\vs}{X\paren{\v t}}$ between random variates $X\paren\vs$,
$X\paren{\v t}$ be specified by the isotropic Mat\'{e}rn covariance function,
defined via
\begin{align}
  \kappa\paren{\vs, \v t; \sigma^2, \rho, \nu} =
    \frac{\sigma^2}{2^{\nu-1}\Gamma\paren\nu}
    \paren{\norm{\vs - \v t} / \rho}^\nu
    K_\nu \paren{\norm{\vs - \v t} / \rho},
\end{align}
in which $\norm{\v\cdot}$ is the Euclidean norm, $K_\nu$ is the modified Bessel
function of the second kind with order $\nu>0$, which governs the smoothness of
the process; $\sigma^2>0$ is a scaling parameter; and $\rho>0$ is a range
parameter.  Gaussian processes imply the vector of observations
$\v X = \paren{X\paren{\vs_1}, \dots, X\paren{\vs_N}}^T \in\mathbb R^N$
at locations $\DS = \brc{\vs_1,\dots,\vs_N} \subset \D$ is normally distributed
$\v X \thicksim \mathcal N \paren{\v 0, \Sigma}$.  The covariance matrix
$\Sigma\in\mathbb R^{N\times N}$ is spatially-structured,
with entries $\Sigma_{ij} = \kappa\paren{\vs_i, \vs_j; \sigma^2, \rho, \nu}$.
The Gaussian process assumption allows estimation of the field
$\brc{X\paren\vs}_{\vs\in\D}$ at unobserved locations
$\DS_0 = \brc{\vs_{01}, \dots, \vs_{0M}} \subset \D$ via kriging, which uses
conditional normal distributions for the unobserved responses.
Standard Bayesian hierarchical modeling techniques for spatial data
\citep[e.g.,][Chapter 6]{Banerjee2015} use conjugate or weakly informative
priors for the covariance parameters, specified via
\begin{align}
  \sigma^2 \thicksim& \text{Inverse-Gamma}\paren{a,b}, \\
  \rho \thicksim& \text{Uniform}\paren{L_0,U_0}, \\
  \nu \thicksim& \text{Uniform}\paren{L_1, U_1}.
\end{align}

We simulate one dataset with $N=300$ locations, sampled uniformly from the unit
square $\mathcal D = \brk{0,1}^2$ and with covariance parameters
$\paren{\sigma^2, \rho, \nu} = \paren{1, .3, .5}$.  We then estimate the
covariance parameters as well as the field $\brc{X\paren\vs}_{\vs\in\D}$ at
$M = 400$ unobserved, gridded locations $\DS_0 \subset \D$.  The priors are
specified via $\paren{a,b,L_0,U_0,L_1,U_1} = \paren{2, 1, 0, 1, 0, 1}$.

\subsubsection{Posterior inference and results}

Standard techniques approximate posterior distributions with a Gibbs
sampler and composition sampling \citep[e.g.,][Chapter 6]{Banerjee2015}.
Conjugate distributions are
used to sample the scale $\sigma^2$ and unobserved field values
$\v X_0 = \paren{X\paren{\vs_{01}}, \dots, X\paren{\vs_{0M}}} \in\mathbb R^M$,
but Metropolis steps are used for the range $\rho$ and smoothness $\nu$
parameters.  The Gibbs sampler is used to draw 60,000 posterior samples for
the covariance parameters, taking 2,043 seconds to complete; posterior inference
uses the final 30,000 iterations.  After drawing posterior samples for the
covariance parameters, composition sampling is used to draw samples for the
unobserved field values $\v X_0$ in parallel, taking 608 seconds to
complete \citep[pg. 126]{Banerjee2015}.

We use the BISQuE strategy to approximate the posterior density
$f\paren{\given{\v X_0}{\v X}}$.  Sparse grid quadrature techniques are used to
directly approximate the marginal posterior densities
$f\paren{\given{\sigma^2}{\v X}}$, $f\paren{\given{\rho}{\v X}}$, and
$f\paren{\given{\nu}{\v X}}$.  \Cref{table:fur_seal_mapping} connects this
example's notation to that used with BISQuE.  When used as the BISQuE
conditioning variable $\v\theta_2$, we map covariance parameters to the real
line by log-transforming the scale parameter $\sigma^2$, and logit-transforming
the range $\rho$ and smoothness $\nu$ parameters.
All conditional and marginal posterior densities required for BISQuE are
computable in closed form up to a proportionality constant; refer to
\citet[eqs. 2.15--16]{Banerjee2015} for details.

Posterior inference via BISQuE and sparse grid quadrature is effectively
identical to posterior inference via Gibbs sampling, but is computed with
substantially less effort. Drawing posterior covariance parameter samples takes
2,043 seconds and composition sampling takes an additional 608 seconds, whereas
the BISQuE and sparse grid quadrature approximations take a total of 238 seconds
(\Cref{table:fur_seal_mapping}), and posterior inference is nearly identical
(\Crefrange{fig:plot_MCMC_krig_summary}{fig:plot_post_covs}).

\begin{knitrout}
\definecolor{shadecolor}{rgb}{0.969, 0.969, 0.969}\color{fgcolor}\begin{figure}

{\centering \includegraphics[width=\maxwidth]{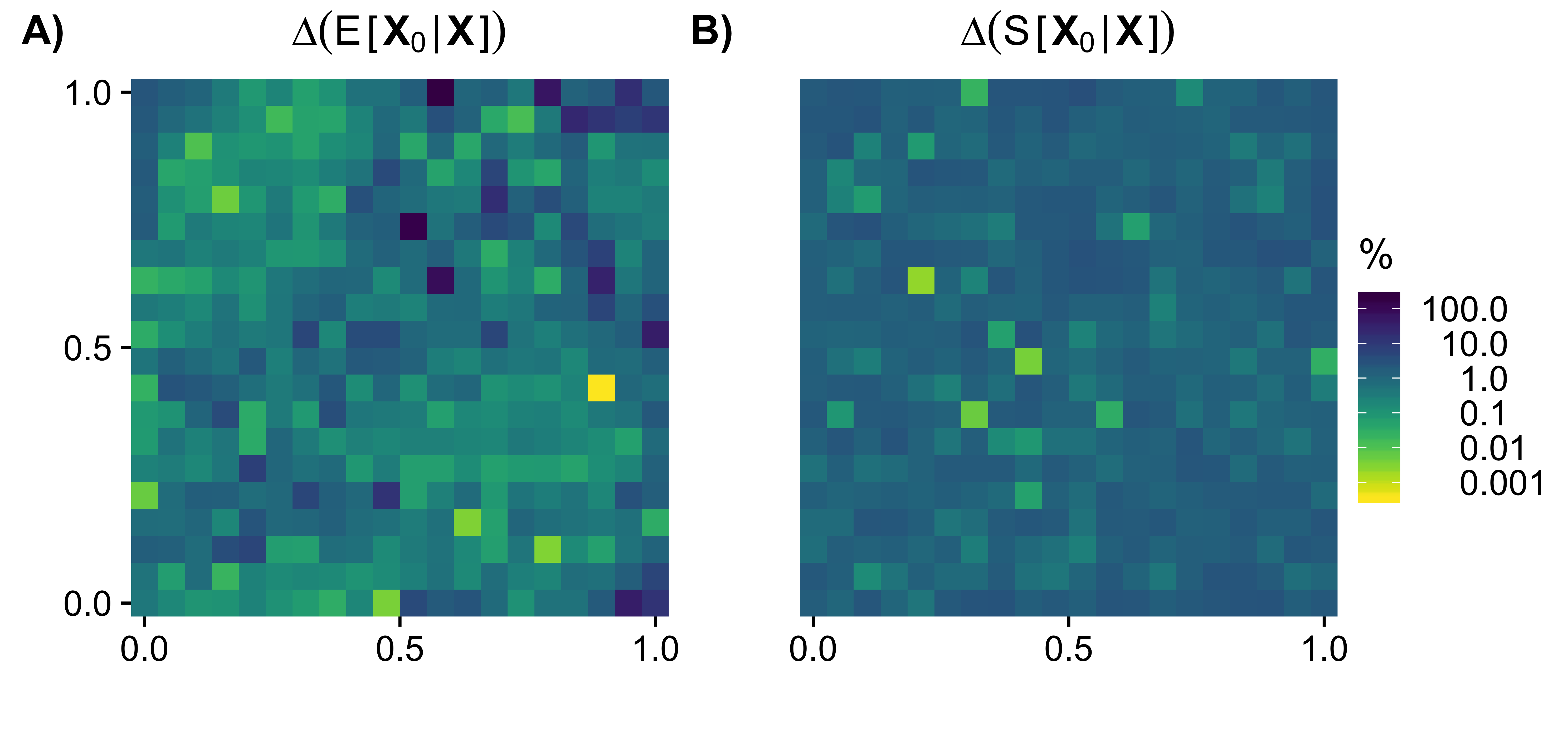} 

}

\caption[Spatial model example]{Spatial model example: Relative differences between BISQuE and Gibbs approximations $\Delta\paren{Y} = \paren{Y_{BISQuE} - Y_{Gibbs}}/Y_{Gibbs} \times 100\%$ for the posterior predictive means (A) and standard errors (B) for the field $\brc{X\paren\vs}_{\vs\in\D}$ at unobserved locations $\DS_0$.  Nearly all (95\%) relative differences in the posterior mean (A) are less than 5.5\% (median=0.4\%); relative differences in the mean are artificially large in regions where the posterior mean is near 0.  All relative differences in the posterior standard errors (B) are below 3.3\% (median=1.4\%).}\label{fig:plot_MCMC_krig_summary}
\end{figure}

\end{knitrout}

\begin{knitrout}
\definecolor{shadecolor}{rgb}{0.969, 0.969, 0.969}\color{fgcolor}\begin{figure}

{\centering \includegraphics[width=\maxwidth]{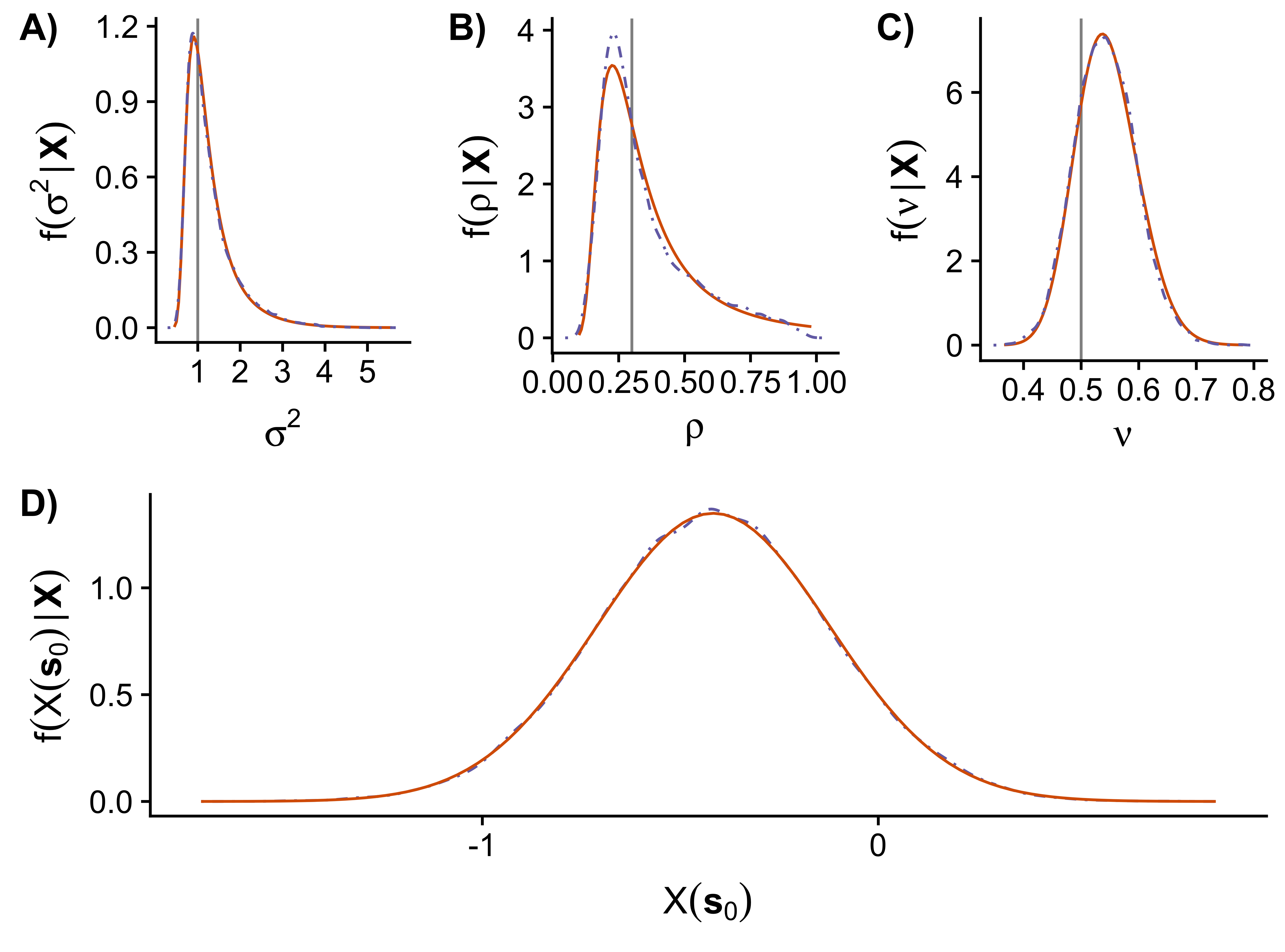} 

}

\caption{Spatial model example: A, B, C) Sparse grid quadrature ({\color{dkrd}---}) and Gibbs ({\color{dkpu}-$\cdot$-}) approximations to the posterior densities for the spatial covariance parameters $\paren{\sigma^2,\rho,\nu}$ are nearly identitcal.  The true values of the parameters are marked by grey vertical lines. D) BISQuE ({\color{dkrd}---}) and Gibbs ({\color{dkpu}-$\cdot$-}) approximations to the posterior density for new observation $X\paren{\v s_0}$ is nearly identical at $\vs_0 = \paren{.5, .2}$, for example.}\label{fig:plot_post_covs}
\end{figure}

\end{knitrout}

\subsection{Remote effects spatial process models}
\label{sec:resp_ex}

\subsubsection{Data and model}

While most spatial data can be modeled with the assumption that distant points
are uncorrelated, large-scale atmospheric circulations can induce dependence
between fields separated by large distances.  The resulting climate
phenomena, known as teleconnection, may be modeled using remote effects
spatial process (RESP) models, which can improve  teleconnection-based
predictions of seasonal precipitation \citep{Hewitt2018}. The RESP model is
given by
\begin{align}
\label{eq:resp_base}
  Y\paren{\vs, t} = \v x^T\paren{\vs ,t}\v\beta + w\paren{\vs, t} +
    \gamma\paren{\vs,t},
\end{align}
which uses a stochastic teleconnection term
\begin{align}
\label{eq:resp_teleconnection}
  \gamma\paren{\vs,t} = \int_{\mathcal D_Z} z\paren{\v r, t}
    \alpha\paren{\vs, \v r} d\v r
\end{align}
 to extend standard geostatistical regression models for a process
$\brc{Y\paren{\vs, t} : \vs \in \mathcal D_Y, t \in \mathcal T}$ defined on a
continuous spatial domain $\mathcal D_Y$ for discrete times $\mathcal T$.
Regression coefficients $\v\beta$ and spatially-correlated variation
$w\paren{\vs ,t}$ are augmented by \autoref{eq:resp_teleconnection}, which
uses doubly-indexed random effects $\alpha\paren{\vs, \v r}$ to
aggregate the impact of remote covariates
$\brc{z\paren{\v r, t} : \v r \in \mathcal D_Z, t\in \mathcal T}$,
such as sea surface temperatures, on a distant response, such as the
standardized deviation $Y\paren{\vs, t}$ from mean seasonal precipitation.
The authors adopt the climate science
convention that mean precipitation is treated as known, and the standardized
deviation $Y\paren{\vs, t}$ is the scientifically interesting response variable
to model.

The RESP model uses two isotropic Mat\'{e}rn covariance functions
$\kappa\paren{\vs, \vs'; \sigma^2_w, \rho_w, \nu_w}$,
$\kappa\paren{\v r, \v r'; \sigma^2_\alpha, \rho_\alpha, \nu_\alpha}$, and a
nugget effect $\sigma^2_\varepsilon$
to define Gaussian processes that model the spatial variation
$\brc{w\paren{\vs, t} : \vs\in\mathcal D_Y}$ and teleconnection effects
$\brc{\alpha\paren{\vs, \v r} : \vs\in\mathcal D_Y, \v r\in\mathcal D_Z}$.
The Mat\'{e}rn smoothness parameters $\nu_w$ and $\nu_\alpha$ are treated as
fixed, and standard priors are used to model the remaining regression
coefficients $\v\beta$ and covariance parameters $\sigma^2_w$, $\rho_w$,
$\sigma^2_\varepsilon$, $\sigma^2_\alpha$, and $\rho_\alpha$
(cf. \Cref{sec:spatial_ex}).

We follow \citet{Hewitt2018} and use the RESP model to analyze Colorado
precipitation data in a statistical downscaling-like scenario.  The RESP model
regresses standardized deviations $Y\paren{\vs ,t}$ from mean
Colorado precipitation observed at $240$ locations $\vs\in\mathcal D_Y$ onto
local surface temperatures $\v x\paren{\vs ,t}$ and
Pacific Ocean sea surface temperatures $z\paren{\v r, t}$.  The model is fit to
Winter averages from 1981--2012 and an ordinal response
$\tilde Y\paren{\vs ,t}\in\brc{v_1,\dots,v_m}$ is predicted for Winter 2013,
given the covariate values $\v x\paren{\vs , t}$ and $z\paren{\v r, t}$ for
$t=2013$.  The distribution for $\tilde Y\paren{\vs ,t}$ is induced by known
cut points $c_0\paren\vs,\dots,c_m\paren\vs$ and defined such that
$P({\tilde Y\paren{\vs,t}= v_i}) =
  P\paren{ c_{i-1}\paren\vs < Y\paren{\vs,t} < c_{i}\paren\vs}$.
In this application, the ordinal response $\tilde Y\paren{\vs ,t}$ represents
below average $v_1$, about average $v_2$, or $v_3$ above average precipitation.

\subsubsection{Posterior inference and results}

\citet{Hewitt2018} construct a Gibbs sampler that approximates posterior
distributions for the RESP model \autoref{eq:resp_base}.  Gibbs sampling is
computationally expensive for the RESP model because two spatially-structured
covariance matrices must be decomposed at each Gibbs iteration.  Let $\v Y$
denote all observations $Y\paren{\vs,t}$ from $t=1981,\dots,2012$; $\v Y_0$
denote all unobserved responses $Y\paren{\vs,t}$ at $t=2013$; and
$\tilde{\v Y}_0$ denote all unobserved ordinal responses $\tilde Y\paren{\vs,t}$
at $t=2013$. Conjugate distributions are used to sample the regression
parameters $\v\beta$, scales $\sigma^2_w$ and $\sigma^2_\alpha$, and continuous
predictions $\v Y_0$; and Metropolis steps are used for the ranges $\rho_w$ and
$\rho_\alpha$.  The Gibbs sampler is used to draw 41,000 posterior samples for
the regression and covariance parameters, taking 8,331 seconds to complete;
posterior inference discards the first 1,000 iterations as the chain mixes
quickly, but requires many iterations to control Monte Carlo integration error.
Composition sampling is then used to draw samples for the predicted response
$\v Y_0$ in parallel, taking 755 seconds to complete. The continuous posterior predictive density $f\paren{\given{\v Y_0}{\v Y}}$ is discretized after
sampling to approximate $f({{\tilde{\v Y}_0}\vert{\v Y}})$ by using the
empirical quantiles of historical precipitation as cut points
$c_0\paren\vs,\dots, c_3\paren\vs$.

We use the BISQuE strategy to approximate the posterior predictive densities
$f\paren{\given{\v Y_0}{\v Y}}$ and $f({{\tilde{\v Y}_0}\vert{\v Y}})$. In
particular, we use the BISQuE strategy to directly approximate
$f({{\tilde{\v Y}_0}\vert{\v Y}})$ by letting
$h\paren{\v\theta_1,\v\theta_2;\v X}$ in \autoref{eq:baseapprox} be the
conditional cumulative
distribution function for $\v Y_0$.  \Cref{table:fur_seal_mapping} connects this
example's notation to that used with BISQuE.  When used as the BISQuE
conditioning variable $\v\theta_2$, we map covariance parameters to the real
line by log-transforming scale parameters $\sigma^2$ and logit-transforming
range parameters $\rho$.  All conditional and marginal posterior densities
required for BISQuE are computable in closed form up to a proportionality
constant; refer to \citet{Hewitt2018} for distributional results.

Posterior inference via BISQuE is effectively identical to posterior inference
via Gibbs sampling, but is computed with substantially less effort. Drawing
posterior covariance parameter samples takes 8,331 seconds and composition
sampling takes an additional 755 seconds, whereas the BISQuE approximations
take a total of 118 seconds (\Cref{table:fur_seal_mapping}), and posterior
inference is nearly identical (e.g., \Cref{fig:resp_comparison}).  The
approximate BISQuE and Gibbs posterior masses
$\hat P({{\tilde Y_0\paren{\vs,t} = v_i}\vert{\v Y}})$
agree to at least two decimal places for all 240 locations $\vs\in\mathcal D_Y$
and values $v_1,v_2,v_3$; additional computing effort can further reduce
approximation errors, but offers limited practical benefit because the
discretization is coarse.

\begin{knitrout}
\definecolor{shadecolor}{rgb}{0.969, 0.969, 0.969}\color{fgcolor}\begin{figure}

{\centering \includegraphics[width=\maxwidth]{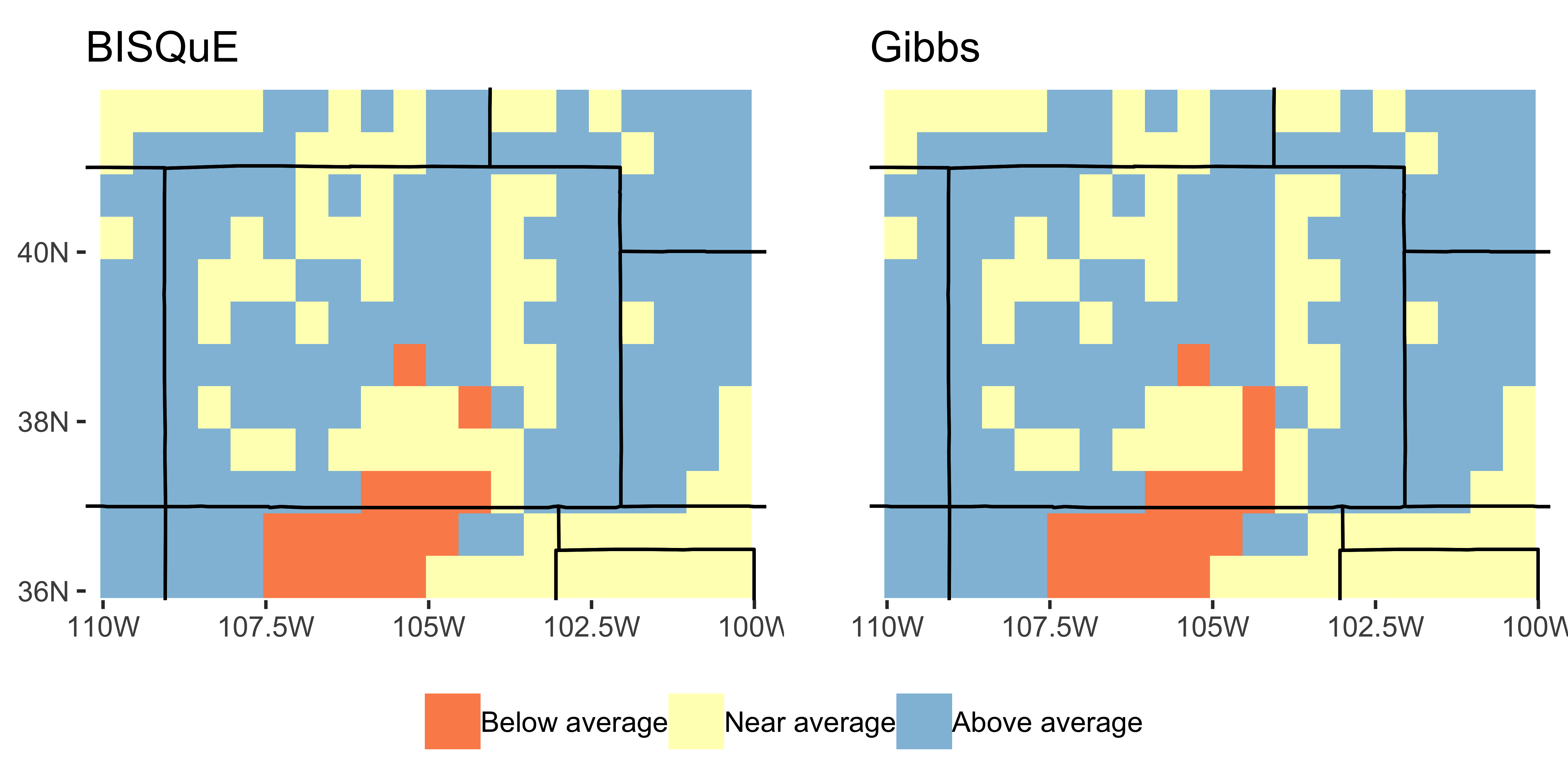} 

}

\caption[BISQuE and Gibbs approximations to the mode of the discretized posterior predictive distributions $f(\tilde{\v Y}_0 \vert \v Y)$ are nearly identical]{BISQuE and Gibbs approximations to the mode of the discretized posterior predictive distributions $f(\tilde{\v Y}_0 \vert \v Y)$ are nearly identical.}\label{fig:resp_comparison}
\end{figure}

\end{knitrout}

\section{Discussion}
\label{sec:discussion}

We combine conditioning techniques with sparse grid quadrature rules to
develop approximate Bayesian Inference via Sparse grid Quadrature
Evaluation (BISQuE).  Approximations \autoref{eq:baseapprox} are developed by
reformulating Bayesian posterior quantities, such as densities and expectations,
so that they may be approximated as weighted mixtures of conditional quantities
$h\paren{\v\theta_1, \v\theta_2; \v X}$.  The integration nodes and weights
from sparse grid quadrature rules are used to build mixing
weights $w^{(i,\ell)}$ and conditioning values $\v\theta_2^{(i,\ell)}$.
In a similar manner as general quadrature techniques and importance sampling
methods, the final BISQuE approximation weights $\tilde w^{(i,\ell)}$ use
weight ratios
${f({{\v\nu^{(i,\ell)}}\vert{\v X}})}/{w({{\v\nu^{(i,\ell)},\v X}})}$
to align the ``theoretical distribution'' $f({\v\nu}\vert{\v X})$ with
the ``sampling distribution'' $w(\v\nu,\v X)$
\citep[pgs. 143, 181]{Givens2013a}.
Nested integration strategies enable BISQuE approximations
\autoref{eq:baseapprox} when models do not have closed form expressions for
required components (\Cref{sec:factored_posteriors}).  Posterior approximation
via BISQuE is deterministic and computationally efficient, offering faster
computation than MCMC methods for a wide range of models \autoref{eq:basemodel}
and posterior quantities \autoref{eq:post_qty}.  In our applications, we find
that BISQuE often reduces overall computing time by at least two orders of
magnitude and yields nearly identical inference to standard MCMC approaches
(\Cref{sec:examples}).

The BISQuE approximation is similar to, and can be
combined with Integrated Nested Laplace approximations (INLA) for latent
Gaussian models \citep{Rue2009a}.  Combining BISQuE with INLA can yield
an approximation technique that scales better to models with more
hyperparameters.  Similar to INLA, our framework will be
most efficient when used to approximate low-dimensional posterior quantities,
like marginal densities or joint densities with computationally tractable
closed form expressions (e.g., $f\paren{\given{\v X_0}{\v X}}$ in
\Cref{sec:spatial_main}).  However, BISQuE does not require that a model have a
latent Gaussian structure and is thus applicable to a broad class of models
such as the population estimation model of \Cref{sec:fur_seals_ex}.

We can combine the BISQuE approximation \autoref{eq:baseapprox} and INLA because
both  methods use conditioning and integration grids to yield fast deterministic
posterior approximation.  In terms of the general hierarchical model
\autoref{eq:basemodel}, INLA specifies a hierarchical parameter model
$f\paren{\v\theta_1, \v\theta_2} =
  f\paren{\given{\v\theta_1}{\v\theta_2}}
  f\paren{\v\theta_2}$
in which $f\paren{\given{\v\theta_1}{\v\theta_2}}$ is Gaussian and
$f\paren{\v\theta_2}$ is a prior distribution for relatively
low-dimensional hyperparameters $\v\theta_2$.  \citet{Rue2009a} define
$\v\theta_1=\paren{\theta_{11},\dots,\theta_{1i},\dots,\theta_{1n}}$,
develop an integration grid, and use Laplace approximations for
$f\paren{\given{\v\theta_{1i}}{\v\theta_2,\v X}}$ and
$f\paren{\given{\v\theta_2}{\v X}}$
to approximate the marginal posterior density
$f\paren{\given{\theta_{1i}}{\v X}}$.  The nested Laplace approximations can be
embedded in the BISQuE approximation \autoref{eq:baseapprox}, yielding posterior
approximation that uses an alternate integration grid to INLA.  The embedding
can be beneficial because sparse grid quadrature rules allow for more
computationally efficient approximation in models with higher dimensional
hyperparameters $\v\theta_2$. Specifically, \citet{Rue2009a} suggest creating
integration grids for models with high-dimensional $\v\theta_2$ by using
central composite design (CCD) methods---an experimental design and response
surface technique for approximating second order surfaces with relatively few
function evaluations \citep{Box1951}.  When integration is the main concern,
sparse grid quadrature methods can require substantially fewer integration
nodes in high dimensions \citep[Table 2, $\ell=3$]{Novak1999} than CCD-based
grids \citep[Table 3]{Sanchez2005}.

Our BISQuE approximation advances Bayesian computing for hierarchical
models, but open questions remain for wider application of the
method.  Notably, our approximation requires the ability to evaluate
$h\paren{\v\theta_1,\v\theta_2;\v X}$ quickly, so may often be limited to
marginal posterior inference for $\v\theta_1$ with relatively small dimension.
Our approximation also relies on the availability of nested quadrature rules for
$\v\theta_2$.  It is difficult to develop quadrature rules for discrete
variables, thus practical use of our approximation may be limited to models with
parameters  $\v\theta_2$ defined on continuous spaces $\Omega_2$.  Fast
convergence of our approximation also relies on the availability of accurate
approximations to $f\paren{\given{\v\theta_2}{\v X}}$.  If the BISQuE
approximation \autoref{eq:baseapprox} has not converged, intuition about
numerical integration suggests the resulting approximation will likely
underestimate posterior variability \citep{Rue2009a}.  However,
\citet[Section 6.5]{Rue2009a} also point out that
$f\paren{\given{\v\theta_2}{\v X}}$ often becomes increasingly Gaussian as the
dimension of $\v\theta_2$ grows since the Bayesian structure will increase
variability and regularity will the dimension, which will help accelerate
convergence.

The BISQuE methodology suggests continued development in several areas.
Additional diagnostics should be developed for wider practical
application of the BISQuE approximation \autoref{eq:baseapprox}.  The
approximation's convergence can be monitored by checking the approximation's
stability as the level $q$ of the underlying sparse grid quadrature rule
\autoref{eq:smolForm} is increased \citep{Laurie1985}.  However, this does not
necessarily provide a diagnostic that can assess how well conditioned a model
\autoref{eq:basemodel} or posterior quantity \autoref{eq:post_qty} is for use
with BISQuE.  Drawing from importance sampling, studying the weight ratio
${f({{\v\nu^{(i,\ell)}}\vert{\v X}})}/{w({{\v\nu^{(i,\ell)},\v X}})}$
in \autoref{eq:baseapprox} at quadrature nodes $\v\nu^{(i,\ell)}$ may help
diagnose practical issues.  Theoretical smoothness properties of
$h\paren{\v\theta_1, \v\theta_2; \v X}$ or concentration of the posterior
density $f\paren{\given{\v\theta_2}{\v X}}$ may also provide insight into the
conditioning for specific models.

Software is available for implementing BISQuE approximations.  We have
developed the \texttt{bisque} package for \texttt{R} that computes BISQuE
approximations for user-specified models.  Custom implementations of BISQuE can
also be developed for specific, high performance applications with the use of
software libraries, including the \texttt{mvQuad} package for \texttt{R}
and the SGMGA libraries for C and C++ \citep{Burkardt2007, Weiser2016}.  These
libraries  contain tables and routines that compute sparse grid quadrature nodes
and weights if $w\paren{\v\nu,\v X}$ is a member
of a standard family of weight functions \citep[Table 5.6]{Givens2013a}.

\section*{Supplementary materials} Additional information and supporting
material for this article is available online at the journal's website.

\section*{Acknowledgements}

This material is based upon work supported by the National Science Foundation
under grant number AGS--1419558. Any opinions, findings, and
conclusions or recommendations expressed in this material are those of the
authors and do not necessarily reflect the views of the National Science
Foundation.

\bibliographystyle{apacite}
\bibliography{references}

\begin{thebibliography}{29}
\expandafter\ifx\csname natexlab\endcsname\relax\def\natexlab#1{#1}\fi
\expandafter\ifx\csname url\endcsname\relax
  \def\url#1{\texttt{#1}}\fi
\expandafter\ifx\csname urlprefix\endcsname\relax\def\urlprefix{URL: }\fi

\bibitem[{Arasaratnam and Haykin(2009)}]{Arasaratnam2009}
Arasaratnam, I. and Haykin, S. (2009) {Cubature Kalman Filters}.
\newblock \textit{IEEE Transactions on Automatic Control}, \textbf{54},
  1254--1269.

\bibitem[{Attias(2000)}]{Attias2000}
Attias, H. (2000) {A Variational Bayesian Framework for Graphical Models}.
\newblock \textit{Advances in neural information processing systems}, 209--215.

\bibitem[{Banerjee et~al.(2015)Banerjee, Carlin and Gelfand}]{Banerjee2015}
Banerjee, S., Carlin, B.~P. and Gelfand, A.~E. (2015) \textit{{Hierarchical
  Modeling and Analysis for Spatial Data}}.
\newblock Boca Raton, FL: CRC Press, second edn.

\bibitem[{Berger(1985)}]{Berger1985}
Berger, J.~O. (1985) \textit{{Statistical Decision Theory and Bayesian
  Analysis}}.
\newblock New York: Springer Science+Business Media, LLC, second edn.

\bibitem[{Box and Wilson(1951)}]{Box1951}
Box, G. E.~P. and Wilson, K.~B. (1951) {On the Experimental Attainment of
  Optimum Conditions}.
\newblock \textit{Journal of the Royal Statistical Society Series B},
  \textbf{13}, 1--45.

\bibitem[{Burkardt(2007)}]{Burkardt2007}
Burkardt, J. (2007) {Sparse Grid Mixed Growth Anisotropic Rules}.
\newblock \textit{https://people.sc.fsu.edu/{\~{}}jburkardt/cpp{\_}src/sgmga}.
\newblock
  \urlprefix\url{https://people.sc.fsu.edu/{~}jburkardt/cpp{\_}src/sgmga/sgmga.html}.

\bibitem[{Emery and Johnson(2012)}]{Emery2012}
Emery, A.~F. and Johnson, K.~C. (2012) {Practical considerations when using
  sparse grids with Bayesian inference for parameter estimation}.
\newblock \textit{Inverse Problems in Science and Engineering}, \textbf{20},
  591--608.

\bibitem[{Gelfand and Smith(1990)}]{Gelfand1990}
Gelfand, A.~E. and Smith, A.~F. (1990) {Sampling-Based Approaches to
  Calculating Marginal Densities}.
\newblock \textit{Journal of the American Statistical Association},
  \textbf{85}, 398--409.

\bibitem[{Genz and Keister(1996)}]{Genz1996}
Genz, A. and Keister, B.~D. (1996) {Fully Symmetric Interpolatory Rules for
  Multiple Integrals over Infinite Regions with Gaussian Weight}.
\newblock \textit{J. Comp. Appl. Math.}, \textbf{71}, 299--309.

\bibitem[{Gerstner and Griebel(1998)}]{Gerstner1998}
Gerstner, T. and Griebel, M. (1998) {Numerical Integration Using Sparse Grids}.
\newblock \textit{Numerical Algorithms}, \textbf{18}, 209--232.

\bibitem[{Givens and Hoeting(2013)}]{Givens2013a}
Givens, G.~H. and Hoeting, J.~A. (2013) \textit{{Computational Statistics}}.
\newblock Hoboken, NJ: John Wiley {\&} Sons, Inc., second edn.

\bibitem[{Heiss and Winschel(2008)}]{Heiss2008}
Heiss, F. and Winschel, V. (2008) {Likelihood approximation by numerical
  integration on sparse grids}.
\newblock \textit{Journal of Econometrics}, \textbf{144}, 62--80.

\bibitem[{Hewitt et~al.(2018)Hewitt, Hoeting, Done and Towler}]{Hewitt2018}
Hewitt, J., Hoeting, J.~A., Done, J.~M. and Towler, E. (2018) {Remote effects
  spatial process models for modeling teleconnections}.
\newblock \textit{Environmetrics}.

\bibitem[{Jia et~al.(2012)Jia, Xin and Cheng}]{Jia2012}
Jia, B., Xin, M. and Cheng, Y. (2012) {Sparse-grid quadrature nonlinear
  filtering}.
\newblock \textit{Automatica}, \textbf{48}, 327--341.

\bibitem[{Laurie(1985)}]{Laurie1985}
Laurie, D.~P. (1985) {Practical error estimation in numerical integration}.
\newblock \textit{Journal of Computational and Applied Mathematics},
  \textbf{12}, 425--431.

\bibitem[{Long et~al.(2013)Long, Scavino, Tempone and Wang}]{Long2013}
Long, Q., Scavino, M., Tempone, R. and Wang, S. (2013) {Fast estimation of
  expected information gains for Bayesian experimental designs based on Laplace
  approximations}.
\newblock \textit{Computer Methods in Applied Mechanics and Engineering},
  \textbf{259}, 24--39.

\bibitem[{Naylor and Smith(1982)}]{Naylor1982}
Naylor, J.~C. and Smith, A. F.~M. (1982) {Applications of a Method for the
  Efficient Computation of Posterior Distributions}.
\newblock \textit{Journal of the Royal Statistical Society, Series C},
  \textbf{31}, 214--225.

\bibitem[{Novak and Ritter(1996)}]{Novak1996}
Novak, E. and Ritter, K. (1996) {High dimensional integration of smooth
  functions over cubes}.
\newblock \textit{Numerische Mathematik}, \textbf{75}, 79--97.

\bibitem[{Novak and Ritter(1999)}]{Novak1999}
--- (1999) {Simple Cubature Formulas with High Polynomial Exactness}.
\newblock \textit{Constructive Approximation}, \textbf{15}, 499--522.

\bibitem[{Rubin(1984)}]{Rubin1984}
Rubin, D.~B. (1984) {Bayesianly Justifiable and Relevant Frequency Calculations
  for the Applied Statistician}.
\newblock \textit{The Annals of Statistics}, \textbf{12}, 1151--1172.

\bibitem[{Rue et~al.(2009)Rue, Martino and Chopin}]{Rue2009a}
Rue, H., Martino, S. and Chopin, N. (2009) {Approximate Bayesian Inference for
  Latent Gaussian Models by Using Integrated Nested Laplace Approximations}.
\newblock \textit{Journal of the Royal Statistical Society Series B},
  \textbf{71}, 319--392.

\bibitem[{Sanchez and Sanchez(2005)}]{Sanchez2005}
Sanchez, S.~M. and Sanchez, P.~J. (2005) {Very large fractional factorials and
  central composite designs}.
\newblock \textit{ACM Transactions on Modeling and Computer Simulation},
  \textbf{15}, 362--377.

\bibitem[{Schillings and Schwab(2013)}]{Schillings2013}
Schillings, C. and Schwab, C. (2013) {Sparse, adaptive Smolyak quadratures for
  Bayesian inverse problems}.
\newblock \textit{Inverse Problems}, \textbf{29}.

\bibitem[{Smolyak(1963)}]{Smolyak1963}
Smolyak, S.~A. (1963) {Quadrature and interpolation formulas for tensor
  products of certain classes of functions}.
\newblock \textit{Doklady Akademii Nauk SSSR}, \textbf{148}, 1042--1045.

\bibitem[{Spiegelhalter et~al.(2002)Spiegelhalter, Best, Carlin and van~der
  Linde}]{Spiegelhalter2002}
Spiegelhalter, D.~J., Best, N.~G., Carlin, B.~P. and van~der Linde, A. (2002)
  {Bayesian measures of model complexity and fit}.
\newblock \textit{Journal of the Royal Statistical Society Series B},
  \textbf{64}, 583--639.

\bibitem[{Tavare et~al.(1997)Tavare, Balding, Griffiths and
  Donnelly}]{Tavare1997}
Tavare, S., Balding, D.~J., Griffiths, R.~C. and Donnelly, P. (1997) {Inferring
  Coalescence Times From DNA Sequence Data}.
\newblock \textit{Genetics}, \textbf{145}, 505--518.

\bibitem[{Tierney and Kadane(1986)}]{Tierney1986}
Tierney, L. and Kadane, J.~B. (1986) {Accurate Approximations for Posterior
  Moments and Marginal Densities}.
\newblock \textit{Journal of the American Statistical Association},
  \textbf{81}, 82--86.

\bibitem[{Watanabe(2010)}]{Watanabe2010}
Watanabe, S. (2010) {Asymptotic Equivalence of Bayes Cross Validation and
  Widely Applicable Information Criterion in Singular Learning Theory}.
\newblock \textit{Journal of Machine Learning Research}, \textbf{11},
  3571--3594.

\bibitem[{Weiser(2016)}]{Weiser2016}
Weiser, C. (2016) {mvQuad: Methods for Multivariate Quadrature}.
\newblock \textit{http://cran.r-project.org/package=mvQuad}.
\newblock \urlprefix\url{http://cran.r-project.org/package=mvQuad}.

\end{thebibliography}


\begin{thebibliography}{}

\bibitem [\protect \citeauthoryear {%
Spiegelhalter%
, Best%
, Carlin%
\BCBL {}\ \BBA {} van~der Linde%
}{%
Spiegelhalter%
\ \protect \BOthers {.}}{%
{\protect \APACyear {2002}}%
}]{%
Spiegelhalter2002}
\APACinsertmetastar {%
Spiegelhalter2002}%
\begin{APACrefauthors}%
Spiegelhalter, D\BPBI J.%
, Best, N\BPBI G.%
, Carlin, B\BPBI P.%
\BCBL {}\ \BBA {} van~der Linde, A.%
\end{APACrefauthors}%
\unskip\
\newblock
\APACrefYearMonthDay{2002}{}{}.
\newblock
{\BBOQ}\APACrefatitle {{Bayesian measures of model complexity and fit}}
  {{Bayesian measures of model complexity and fit}}.{\BBCQ}
\newblock
\APACjournalVolNumPages{Journal of the Royal Statistical Society Series
  B}{64}{4}{583--639}.
\PrintBackRefs{\CurrentBib}

\bibitem [\protect \citeauthoryear {%
Watanabe%
}{%
Watanabe%
}{%
{\protect \APACyear {2010}}%
}]{%
Watanabe2010}
\APACinsertmetastar {%
Watanabe2010}%
\begin{APACrefauthors}%
Watanabe, S.%
\end{APACrefauthors}%
\unskip\
\newblock
\APACrefYearMonthDay{2010}{}{}.
\newblock
{\BBOQ}\APACrefatitle {{Asymptotic Equivalence of Bayes Cross Validation and
  Widely Applicable Information Criterion in Singular Learning Theory}}
  {{Asymptotic Equivalence of Bayes Cross Validation and Widely Applicable
  Information Criterion in Singular Learning Theory}}.{\BBCQ}
\newblock
\APACjournalVolNumPages{Journal of Machine Learning
  Research}{11}{}{3571--3594}.
\newblock
\begin{APACrefURL}
  \url{http://www.jmlr.org/papers/volume11/watanabe10a/watanabe10a.pdf}
  \end{APACrefURL}
\PrintBackRefs{\CurrentBib}

\end{thebibliography}

\end{document}